\documentclass[prd,showpacs]{revtex4}
\usepackage{graphicx}
\usepackage{bm}

\newcommand{\Tr}{{\rm Tr}}
\newcommand{\bfl}{{\bf l}}
\newcommand{\nhat}{{\hat{\bf n}}}

\newcommand{\aap}{{A\&A}}
\newcommand{\araa}{{ARA\&A}}
\newcommand{\mnras}{{MNRAS}}

\begin{document}

\title{Reconstruction of lensing from the cosmic microwave background
polarization}
\author{Christopher M. Hirata\footnote{chirata@princeton.edu} \& Uro\v s
Seljak\footnote{useljak@princeton.edu}}
\affiliation{Department of Physics, Jadwin Hall, Princeton University,
Princeton NJ 08544, USA}
\date{July 27, 2003}

\begin{abstract} 
Gravitational lensing of the cosmic microwave background (CMB)
polarization field has been recognized as a potentially valuable probe of
the cosmological density field.  We apply likelihood-based techniques to
the problem of lensing of CMB polarization and show that if the $B$-mode
polarization is mapped, then likelihood-based techniques allow
significantly better lensing reconstruction than is possible using the
previous quadratic estimator approach.  With this method the ultimate
limit to lensing reconstruction is not set by the lensed CMB power
spectrum. Second-order corrections are known to produce a curl component
of the lensing deflection field that cannot be described by a potential;
we show that this does not significantly affect the reconstruction at
noise levels greater than 0.25 $\mu$K arcmin.  The reduction of the mean
squared error in the lensing reconstruction relative to the quadratic
method can be as much as a factor of two at noise levels of 1.4 $\mu$K
arcmin to a factor of ten at 0.25 $\mu$K arcmin, depending on the angular
scale of interest.
\end{abstract}

\maketitle

\section{Introduction}

Over the past decade the cosmic microwave background (CMB) anisotropy has
been established as a robust and powerful cosmological probe.  While much
attention has focused on the primary anisotropy generated in the early
universe, the CMB should also contain signatures of processes that
occurred between the surface of last scatter and the present.  One of
these is weak gravitational lensing, which has been recognized as a probe
of large scale structure (LSS)
\cite{1997A&A...324...15B,1999PhRvD..59l3507Z, 1999PhRvL..82.2636S,
2001ApJ...557L..79H, 2001PhRvD..63d3501B}. Aside from its use as a probe
of the matter power spectrum at low redshift $z\ll 1100$, weak lensing of
the CMB could be cross correlated against other tracers of the density
field such as galaxy surveys \cite{2000ApJ...540..605P} or weak lensing of
galaxies \cite{2001PhRvD..63d3501B}.  Through cross-correlation with the
CMB temperature, an improved measurement of the integrated Sachs-Wolfe
effect over that possible using the CMB power spectrum alone is possible,
yielding constraints on the late-time growth function and hence on the
dark energy \cite{1999PhRvD..59j3002G, 1999PhRvD..60d3504S}. Lensing has
also attracted attention recently as a cosmological source of $B$-mode
polarization \cite{1998PhRvD..58b3003Z}; reconstruction and removal of
lensing $B$-modes will thus be an important part of a future search for
$B$-mode polarization induced by primordial gravitational waves
\cite{2002ApJ...574..566H, 2002PhRvL..89a1303K, 2002PhRvL..89a1304K}.

The lensing signal in the CMB is small, so it is important to construct
optimal methods for estimating the lensing field from CMB data. The early
investigations of lensing of the CMB temperature showed that while there
is an effect of lensing on the CMB power spectrum
\cite{1989MNRAS.239..195C,1996ApJ...463....1S}, it is much more promising
to estimate the lensing field using quadratic combinations of the CMB
temperature, and to estimate the lensing power spectrum using the
four-point correlation function (or its harmonic equivalent, the
trispectrum) \cite{1997A&A...324...15B, 1999PhRvL..82.2636S,
2000PhRvD..620a9fdG}. More recent work has identified the divergence of
the temperature-weighted gradient as the optimal quadratic combination of
the CMB temperature for use in lensing studies
\cite{2000PhRvD..62f3510Z,2001ApJ...557L..79H}. Analysis based on
likelihood techniques \cite{2003PhRvD..67d3001H} has since shown that the
quadratic estimator technique is statistically optimal when the lensing
effect on the CMB covariance matrix is small.  This was shown to be a good
approximation for lensing of temperature anisotropies in the range $l\le
3500$. For the small scales $l\gg 3500$, the primary CMB power spectrum is
much smaller than the lensed power spectrum, hence this argument breaks
down. In this case for sufficiently small instrument noise the
reconstruction of projected mass density can be nearly perfect
\cite{2000ApJ...538...57S}. We will not discuss the reconstruction on
these very small scales in this paper.

Our ability to reconstruct the lensing field using the CMB temperature is
limited because the temperature fluctuations are stochastic and so we can
only statistically determine the unlensed CMB temperature field. It is
thus advantageous to consider lensing of the CMB polarization, since in
the absence of primordial gravitational waves the unlensed CMB
polarization is entirely in $E$ rather than $B$ modes.  This implies that,
in the terminology of galaxy lensing, there is no ``shape noise'' in the
CMB polarization field. Several authors have developed algorithms that use
the $B$-modes induced by lensing to probe LSS \cite{2000PhRvD..62d3007H,
2001PhRvD..63d3501B}.  The optimal quadratic estimator -- the polarization
analogue of the temperature-based quadratic estimator using the divergence
of the temperature-weighted gradient -- was constructed by Ref.
\cite{2002ApJ...574..566H}. There it was shown that for sufficiently small
detector noise most of the lensing reconstruction information with this
method is provided by the $B$ mode.

Even with polarization information these quadratic estimators cannot
improve the reconstruction beyond a certain level, set by the coherence
length of the polarization. It has been argued that this provides a
fundamental limit to our ability to separate the lensing induced $B$ modes
from the $B$ modes induced by gravity waves
\cite{2002PhRvL..89a1303K,2002PhRvL..89a1304K}.  However, it has not been
determined whether quadratic estimation is optimal for the
polarization-based lensing reconstruction, and indeed Refs.
\cite{2002PhRvL..89a1303K,2002PhRvL..89a1304K} comment that it might be
possible to extract additional information in higher-order statistics.  
The argument for optimality of the quadratic estimator presented by Ref.
\cite{2003PhRvD..67d3001H} does not apply to polarization since the
$B$-mode power is dramatically increased by lensing. Here we construct
likelihood-based estimators for lensing using the CMB polarization and
show that the likelihood-based estimator improves significantly on the
quadratic estimator (although we do not present these as series of
higher-order statistics).  Indeed, as noise is decreased the accuracy of
CMB lensing reconstruction continues to improve without bound.  
Conceptually this is because if the lensed polarization is measured with
zero noise, then the equation $B_{\rm unlensed}=0$ can be solved (except
possibly for a small number of degenerate modes) for the projected matter
density with zero noise.  The equation $B_{\rm unlesed}=0$ is ill-behaved
in the presence of instrument noise; fortunately, the likelihood formalism
easily incorporates noise and, as we show in this paper, regularizes the
problem.

In practice, a perfect reconstruction of the lensing potential is
impossible because as instrument noise is reduced, some contaminant to the
lensing signal will eventually become more important than the instrument
noise.  One candidate for this limiting factor is lensing field rotation
caused by the fact that the density perturbations causing the lensing are
spread out along the line of sight (i.e. there is more than one ``lens
plane'') and that the lensing is not perfectly weak (i.e. the first-order
Born approximation to the lensing field is inexact). We will show that
even for an experiment with noise 0.25 $\mu$K arcmin and 2 arcmin
full-width half maximum beam, the field rotation does not substantially
worsen the lensing reconstruction. It is however possible that foreground
contamination will be a more serious problem.

Studies of lensing of CMB polarization will require that the polarization
field be mapped with noise levels of the order of $\sim 1$ $\mu$K arcmin;
this would be a substantial improvement in sensitivity beyond that of the
current {\slshape Wilkinson Microwave Anisotropy Probe} ({\slshape WMAP};
{\slshape http://map.gsfc.nasa.gov/}) and the upcoming {\slshape Planck}
({\slshape http://astro.estec.esa.nl/Planck/}) experiments (see Table
\ref{tab:expt}).  Nevertheless, $\sim 1$ $\mu$K arcmin may be achieved
with a future polarization satellite.  The noise levels of $<$0.25 $\mu$K
arcmin at which field rotation becomes important will probably remain
unachievable for the foreseeable future.

This paper is organized as follows: in \S\ref{sec:formalism}, we define
our notations and conventions. In \S\ref{sec:likelihood}, we consider the
properties of the likelihood function and its implications for likelihood
and Bayesian analyses of CMB lensing reconstruction. In
\S\ref{sec:crosspotential}, we investigate the breakdown of the Born
approximation for CMB lensing and its implications for lensing
reconstruction. Determination of the lensing power spectrum from CMB maps
is discussed in \S\ref{sec:ape}. We show numerical simulations of CMB
polarization lensing and reconstruction using our estimators in
\S\ref{sec:numer}, and conclude in \S\ref{sec:conclusions}.

The fiducial cosmology used in these simulations is a spatially flat
cosmological constant-dominated universe with baryon fraction $\Omega_{b0}
= 0.046$; cold dark matter fraction $\Omega_{c0} = 0.224$; cosmological
constant fraction $\Omega_{\Lambda 0} = 0.73$; Hubble parameter $H_0=72$
km/s/Mpc; primordial helium abundance $Y_P=0.24$; reionization optical
depth $\tau_r = 0.17$; primordial scalar spectral index $n_s=1$; and no
primordial vector or tensor perturbations. We have used the {\sc cmbfast}
numerical package \cite{1996ApJ...469..437S} to compute all power spectra
except in \S\ref{sec:effect}. The experiments considered are as shown in
Table \ref{tab:expt}.  The {\slshape WMAP} and the {\slshape Planck} will
not be able to map the lensing field using polarization and are included
in the table for comparison.  The Reference Experiments A through F are
successively lower-noise (or finer-beam) experiments that were analyzed to
determine how the signal-to-noise ratio in the lensing reconstruction
depends on experimental parameters.  Note that Experiment C is the
Reference Experiment of Refs.
\cite{2002ApJ...574..566H,2002PhRvL..89a1303K}.

\begin{table}
\caption{\label{tab:expt}Parameters for CMB experiments.}
\begin{tabular}{lcccccccc}
\hline\hline
Experiment & & ${\cal N}_P$ / $\mu$K arcmin & & $\theta_{FWHM}$ / arcmin
& & $l_c$  \\
\hline
{\slshape WMAP}, 4 yrs. (94 GHz) & & 700 & & 13 & & 620 \\
{\slshape Planck}, 1 yr. (143 GHz) & & 81 & & 8 & & 1010 \\
Ref. Expt. A & & 3.0 & & 7 & & 1160 \\
Ref. Expt. B & & 1.41 & & 7 & & 1160 \\
Ref. Expt. C & & 1.41 & & 4 & & 2020 \\
Ref. Expt. D & & 1.00 & & 4 & & 2020 \\
Ref. Expt. E & & 0.50 & & 2 & & 4050 \\
Ref. Expt. F & & 0.25 & & 2 & & 4050 \\
\hline\hline
\end{tabular}
\end{table}

\section{Formalism}
\label{sec:formalism}

Here we describe our normalization conventions; note that for some
quantities, there are many conventions in use in the literature, and
appropriate conversion factors must be applied if one wishes to compare
results.

\subsection{CMB}

We work in the normalized flat-sky approximation, i.e. the sky is taken to
be a flat square of side length $\sqrt{4\pi}$ (i.e. total area $4\pi$)
with periodic boundary conditions.  The CMB temperature and polarization
fields can then be expressed as a sum over Fourier modes:
\begin{equation}
\left( \begin{array}{c}
T(\nhat) \\ Q(\nhat) \\ U(\nhat) \end{array} \right)
 = {1\over\sqrt{4\pi}}\sum_\bfl 
\left( \begin{array}{c}
T_\bfl \\ Q_\bfl \\ U_\bfl \end{array} \right)
e^{i\bfl\cdot\nhat},
\end{equation}
where the $\bfl$-modes are distributed in the two-dimensional $\bfl$-space
with number density $1/\pi$.  Defining the angle of a mode by
$\tan\phi_\bfl = l_y/l_x$, we have $E$ and $B$ polarization modes given
by:
\begin{equation}
\left( \begin{array}{c}
E_\bfl \\ B_\bfl \end{array} \right) =
\left( \begin{array}{cc}
\cos (2\phi_\bfl) & \sin (2\phi_\bfl) \\ -\sin (2\phi_\bfl) & \cos
(2\phi_\bfl) \end{array} \right) 
\left( \begin{array}{c}
Q_\bfl \\ U_\bfl \end{array} \right) .
\label{eq:eb}
\end{equation}
(Technically the angle $\phi_0$ of the $\bfl=0$ mode is undefined, however
this will not concern us since within the flat-sky approximation we will
convert sums over $\bfl$ into integrals: $\sum_\bfl\rightarrow\int
d^2\bfl/\pi$.  If an integral is divergent at ${\bfl}=0$, then it cannot
be computed accurately within the flat-sky approximation.)

We will use the following notations for CMB fields: $\{T,Q,U\}$ for the
unlensed (primary) CMB anisotropies; $\{\tilde T,\tilde Q,\tilde U\}$ for
the lensed CMB anisotropies; and $\{\hat T,\hat Q,\hat U\}$ for the
measured anisotropies (including noise).  These are measured in $\mu$K
(blackbody temperature), and we will assume that the monopole (mean
temperature) and special-relativistic effects (kinematic dipole/quadrupole
and stellar aberration)  have been removed.  The instrument noise will be
assumed to be statistically uncorrelated with any cosmological signal and
will be denoted by $\eta_X$, where $X$ is one of $T$, $Q$, or $U$ (or $T$,
$E$, and $B$ depending on which basis is more convenient).  The unlensed
CMB will have a power spectrum given by:
\begin{equation}
\langle X_\bfl^\ast X'_{\bfl'} \rangle = C_l^{XX'} \delta_{\bfl,\bfl'},
\end{equation}
where here $X$ and $X'$ are $T$, $E$, or $B$ (here we desire rotational
symmetry so we cannot use $Q$ or $U$).  We assume the universe is
statistically parity-invariant so that $C_l^{TB}=C_l^{EB}=0$; in some
parts of this paper we will discuss universes with no tensor
perturbations, in which case we also have $C_l^{BB}=0$.

\begin{figure}
\includegraphics[angle=-90,width=5in]{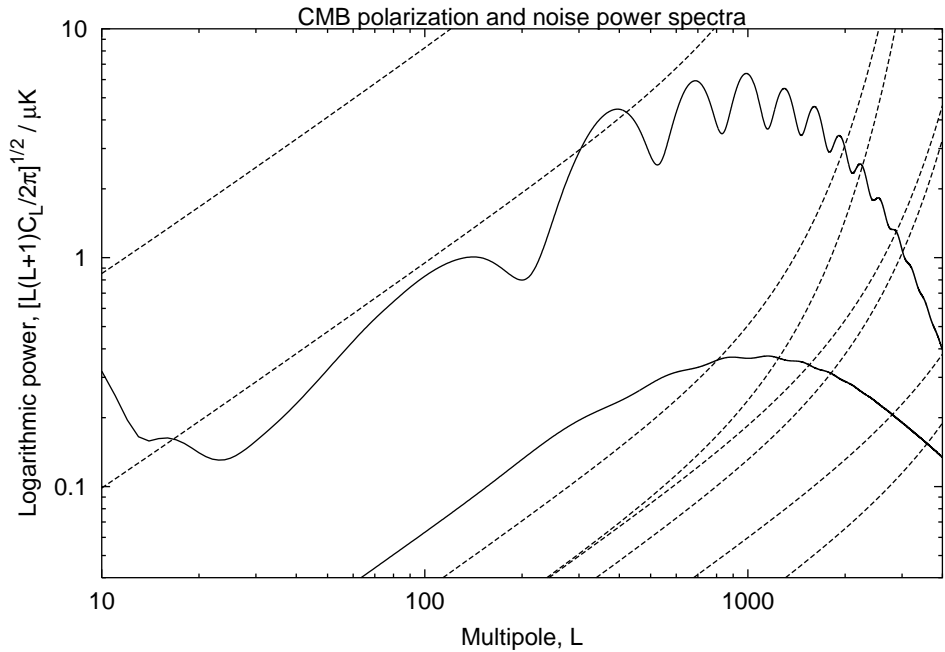}
\caption{\label{fig:cl}CMB polarization power spectra for $E$-type and
$B$-type polarization (upper and lower solid curves, respectively).  The
noise curves for the experiments of Table \ref{tab:expt} are shown as
dashed lines; from top to bottom: {\slshape WMAP}, {\slshape Planck}, and
Reference Experiments A, B, C, D, E, and F. }
\end{figure}

Throughout most of this paper we will take general noise covariance ${\bf
N}$; when we wish to show expected performance for particular experiments,
we will use the following noise power spectrum appropriate for a Gaussian
beam profile:
\begin{equation}
N^{TT}_l = {\cal N}_T^2 e^{l(l+1)\theta_{FWHM}^2/8\ln 2}
= {\cal N}_T^2 e^{l(l+1)/l_c(l_c+1)},
\end{equation}
where $\theta_{FWHM}$ is the full-width at half maximum (FWHM) of the
beam.  We take a similar form for $N^{EE}_l=N^{BB}_l$, except that ${\cal
N}_T$ is replaced with ${\cal N}_P$.  The quantities ${\cal N}_T$, ${\cal
N}_P$, and $\theta_{FWHM}$ (combined with the fraction $f_{sky}$ of the
sky surveyed) thus parameterize the performance of the experiment.  Noise
curves compared to the CMB for the experiments shown in Table
\ref{tab:expt} are shown in Fig. \ref{fig:cl}.  The $l$-value at which the
beam transfer function drops to $1/\sqrt{e}$ is given approximately by
$l_c\approx (8095\,{\rm arcmin})/\theta_{FWHM}$.

We will introduce a vector ${\bf x}$ containing both temperature and
polarization information: ${\bf x} = (T,Q,U)$.  The lensed and measured
temperature/polarization vectors will be denoted $\tilde{\bf x}$ and
$\hat{\bf x}$, respectively.  [Note: ${\bf x}$ is a ``vector'' in the
sense of linear algebra, i.e. it is an element of a vector space, in this
case a Hilbert space with the usual $L^2(S^2)$ inner product, on which
matrix operations such as ${\bf C}$ can act.  It is not a vector in the
sense of differential geometry.] Since most of the fields we deal with,
including CMB temperature and polarization, are real, their Fourier modes
satisfy e.g. $T_\bfl=T_{-\bfl}^\ast$.  Consequently, if we have $N$
Fourier modes, the $N$-dimensional vector with components $\{T_\bfl\}$
only has $N/2$ independent complex components; the remainder contain
redundant information.  (Of course, there are still $N$ independent real
components.) The covariance matrix ${\bf C}$ is defined as ${\bf
C}^{\hat{\bf x}\hat{\bf x}} = \langle \hat{\bf x}\hat{\bf
x}^\dagger\rangle$; note that it is Hermitian by construction.

\subsection{Lensing}

The lensed temperature and polarization are given in terms of the unlensed
temperature by means of the re-mapping function $g$: $\tilde T(\nhat) =
T[g(\nhat)]$, and similarly for $Q$ and $U$ [however, $E$ and $B$ do not
transform this way, rather one must use Eq. (\ref{eq:eb})].  The
re-mapping function can be decomposed into a longitudinal part generated
by the lensing potential $\Phi$, and a transverse part generated by the
lensing cross-potential $\Omega$:
\begin{equation}
g(\nhat) = \nhat + \nabla\Phi(\nhat) + \star\nabla\Omega(\nhat),
\label{eq:g}
\end{equation}
where $\star$ is the ninety-degree rotation operator: $\star\hat{\bf e}_x
= \hat{\bf e}_y$, $\star\hat{\bf e}_y = -\hat{\bf e}_x$. Past studies of
CMB lensing reconstruction have ignored the cross-potential since (for
scalar perturbations) it vanishes at first order in perturbation theory.  
In principle it could become important given the high precision enabled by
lensing of CMB polarization.  However, we will show in
\S\ref{sec:crosspotential} and \S\ref{sec:numer1} that the cross-potential
is unimportant for most near-term experiments.

We restrict our attention to the weak lensing regime, i.e. we assume the
magnification matrix:
\begin{equation}
{\bf M} =
{\partial g(\nhat)\over\partial\nhat} =
\left( \begin{array}{cc} {\partial g_x\over\partial x} & {\partial
g_x\over\partial y} \\
{\partial g_y\over\partial x} & {\partial g_y\over\partial y}
\end{array}\right)
= \left( \begin{array}{cc} 1 + \partial_x^2\Phi -
\partial_x\partial_y\Omega & \partial_x\partial_y\Phi - \partial_y^2\Omega
\\
\partial_x\partial_y\Phi + \partial_x^2\Omega & 1 + \partial_y^2\Phi +
\partial_x\partial_y\Omega \end{array} \right)
= \left( \begin{array}{cc} 1+\kappa+\gamma_Q & \gamma_U+\omega \\
\gamma_U-\omega & 1+\kappa-\gamma_Q \end{array} \right)
\label{eq:weaklens}
\end{equation}
is everywhere invertible.  This is a necessary and sufficient condition to
disallow caustics and multiple images of the same portion of the surface
of last scatter.  Lensing by large-scale structure is too weak to create
caustics on the surface of last scatter, so the weak lensing assumption is
violated only in the vicinity of astrophysical objects such as clusters.  
We classify clusters and other strong lenses as foreground contaminants
and do not consider them further here.  In the regime where the lensing
distortion is small -- i.e. ${\bf M}$ is close to the identity -- we may
interpret the four components $\kappa,\gamma_Q,\gamma_U,\omega$ in Eq.
(\ref{eq:weaklens}) as follows.  The convergence $\kappa$ magnifies a
feature on the last-scattering surface of infinitesimal angular size
$d\vartheta$ to size $(1+\kappa)d\vartheta$.  The field rotation angle
$\omega$ rotates the feature clockwise by $\omega$ radians.  The $Q$-shear
$\gamma_Q$ produces a stretching along the $x$-axis while compressing it
along $y$: the apparent angular extents of a feature along the two axes
are $(1+\gamma_Q)d\vartheta_x$ and $(1-\gamma_Q)d\vartheta_y$,
respectively.  The $U$-shear has a similar effect, stretching along the
$y=x$ axis and compressing along the $y=-x$ axis. It should be noted that
the four fields $\kappa,\gamma_Q,\gamma_U,\omega$ are {\em not}
independent because they are all generated by differentiating the two
fields $\Phi$ and $\Omega$.  In particular, if we $EB$-decompose the shear
field into its positive-parity ($\epsilon$) and negative-parity ($\beta$)
components:
\begin{eqnarray}
\epsilon_\bfl = && 
[\gamma_Q]_\bfl \cos 2\phi_\bfl + [\gamma_U]_\bfl \sin 2\phi_\bfl,
\nonumber \\
\beta_\bfl = && 
-[\gamma_Q]_\bfl \sin 2\phi_\bfl + [\gamma_U]_\bfl \cos 2\phi_\bfl,
\end{eqnarray}
we find that $\epsilon_\bfl=\kappa_\bfl$ and $\beta_\bfl=\omega_\bfl$.  
These are then related to the potentials via $\kappa_\bfl =
(l^2/2)\Phi_\bfl$ and $\omega_\bfl= (l^2/2)\Omega_\bfl$.  This immediately
implies the power spectrum relations $C^{\kappa\kappa}_l =
C^{\epsilon\epsilon}_l= {1\over 4}l^4 C^{\Phi\Phi}_l$ and
$C^{\omega\omega}_l = C^{\beta\beta}_l={1\over 4}l^4 C^{\Omega\Omega}_l$.  
It is of interest to note that the convergence and field rotation can be
determined from the deflection angle ${\bf d}(\nhat) = g(\nhat)-\nhat$ by:
\begin{equation}
\kappa = -{1\over 2} \nabla\cdot{\bf d} \;\;\;{\rm and}\;\;\;
\omega = {1\over 2} \nabla\cdot\star{\bf d}. 
\label{eq:pdec}
\end{equation}

If $\kappa,\gamma_Q,\gamma_U,\omega$ are not small compared to 1, then the
physical interpretation of these quantities is somewhat more complicated.  
We will continue to call $\kappa$ the ``convergence,''
$(\gamma_Q,\gamma_U)$ the ``shear,'' and $\omega$ the ``field rotation
angle'' even in this case, although this convention is not universally
followed in the literature.  Note, however, that the relations
$\epsilon_\bfl=\kappa_\bfl$ and $\beta_\bfl=\omega_\bfl$ continue to hold
(in fact, they remain valid even for strong lenses!), which makes our
definitions of convergence, shear, and field rotation angle particularly
convenient.

Sometimes we will specify a lens re-mapping $g$ by its convergence and
field rotation, $g\equiv (\kappa,\omega)$.  Most authors have performed
the lensing analysis using $\Phi$ rather than $\kappa$ as the field to be
estimated, since the deflection angle is a local function of the former.  
In the present analysis, we take $\kappa$ (and $\omega$ when it is
important) to be the fundamental field.  Of course, the two fields contain
exactly the same information, being related by the differential relation
$\kappa = -{1\over 2}\nabla^2\Phi$ in real space and by a multiplicative
factor of $l^2/2$ in harmonic space.

It is convenient to introduce the lensing operator $\Lambda_g$ defined by
$\Lambda_gX(\nhat) = X(g(\nhat))$, where $X$ is one of $T$, $Q$, or $U$.
[In the $\{T,E,B\}$ basis, the action of $\Lambda_g$ is not so simple and
the transformation of Eq. (\ref{eq:eb}) must be applied.] We define the
$\sigma^\kappa$ differential operator (and analogously $\sigma^\omega$) as
the action of an infinitesimal lens configuration: $\sigma^\kappa_\bfl
\equiv ({\partial\Lambda_g/\partial\kappa_\bfl[g]} )|_{g=(0,0)}$. Then,
noting that $\Lambda_{(0,0)}$ is the identity operator ${\bf 1}$, we find
that to first order in $(\kappa,\omega)$:
\begin{equation}
\Lambda_{(\kappa,\omega)} = 
{\bf 1} + \sum_\bfl \left(\kappa_\bfl \sigma^\kappa_\bfl 
+ \omega_\bfl \sigma^\omega_\bfl\right)
+ O(\kappa^2,\omega^2,\kappa\omega).
\label{eq:lambdaapprox}
\end{equation}
The $\sigma$ operator acts as follows on a field $X$ in the $\{T,Q,U\}$ basis:
\begin{eqnarray}
\left( \sigma^\kappa_\bfl
X \right)_{\bfl'}
= \left( \frac{2}{l^2}\right) {\bfl\cdot(\bfl'-\bfl)\over\sqrt{4\pi}}
X_{\bfl'-\bfl},
\nonumber \\
\left( \sigma^\omega_\bfl
 X \right)_{\bfl'}
= \left(
\frac{2}{l^2}\right) {\bfl\cdot\star(\bfl'-\bfl)\over\sqrt{4\pi}}
X_{\bfl'-\bfl}.
\label{eq:sigma}
\end{eqnarray}
In the $\{T,E,B\}$ basis, the $\sigma$-matrices mix $E$ and $B$ because
these are non-local quantities.  Specifically, they have components:
\begin{equation}
[\sigma^\kappa_\bfl]_{\bfl_1,-\bfl_2} = 
-{\delta_{\bfl_1,\bfl-\bfl_2}\over\sqrt{4\pi}}
\left( \frac{2}{l^2}\right)
(\bfl\cdot\bfl_2)
\left( \begin{array}{ccc}
1 & 0 & 0 \\
0 & \cos 2\alpha & -\sin 2\alpha \\
0 & \sin 2\alpha & \cos 2\alpha
\end{array} \right),
\label{eq:sigmamatrix}
\end{equation}
where the rows correspond to the $\{T,E,B\}_{\bfl_1}$ and the columns to
the $\{T,E,B\}_{-\bfl_2}$, and we have defined the angle $\alpha =
\phi_{\bfl_1}-\phi_{\bfl_2}$.  The matrix for $\sigma^\omega_\bfl$ differs
by replacing the prefactor $\bfl\cdot\bfl_2$ with $\star\bfl\cdot\bfl_2$.
The $\sigma$ matrices satisfy $\sigma^\kappa_\bfl =
(\sigma^\kappa_{-\bfl})^\dagger$.

Lensing alters the CMB anisotropy covariance; the covariance matrix ${\bf
C}^{\tilde X\tilde X'}$, where $X\in\{T,Q,U\}$ (or $X\in\{T,E,B\}$)  of
the lensed temperatures is dependent on the lens configuration $g$, and
thus we will denote it by $\tilde{\bf C}^{XX'}_g \equiv{\bf C}^{\tilde
X\tilde X'}_g$. Since the lensed CMB field is $\tilde{\bf x} =
\Lambda_g{\bf x}$, we have $\tilde{\bf C}_g = \Lambda_g{\bf
C}\Lambda_g^\dagger$. The lensed covariance averaged over the ensemble of
LSS configurations will be denoted by $\langle \tilde{\bf C}^{XX'}
\rangle_{LSS}$.  Note, however, that whereas the primary CMB is expected
to be nearly Gaussian, the lensed CMB is non-Gaussian and so $\langle
\tilde{\bf C}^{XX'} \rangle_{LSS}$ does {\em not} specify completely the
statistics of the lensed CMB field.  Indeed, it is the non-Gaussianity of
the lensed CMB that enables separation of the lensing and gravitational
wave contributions to $B$.  It also means that the standard Gaussian
formula for the uncertainty in the power spectrum,
$\sigma(C_l)/C_l=\sqrt{2/(2l+1)f_{sky}\Delta l}$, does not necessarily
apply to $B$-mode polarization on small scales.

It is readily apparent from Eq. (\ref{eq:sigmamatrix}) that lensing can
produce $B$-modes even if these are not present in the primary CMB.  We
show in Appendix \ref{sec:bmode} that for ``almost all'' primary CMB
realizations, there are only a small number of convergence modes that do
not produce $B$-type polarization.

\subsection{Chi-squared analysis of lensing}
\label{sec:chisq}

We illustrate our formalism with a simple lensing reconstruction via
$\chi^2$ minimization (the ``least squares'' method).  We perform a full
likelihood analysis in \S\ref{sec:likelihood}, but the $\chi^2$ analysis
is sufficiently similar that it illustrates the basic concept.  Define the
functional $\chi^2(\kappa)$ of a convergence field given CMB data
$\hat{\bf x}$:
\begin{equation}
\chi^2(\kappa) \equiv 
(\Lambda_{(\kappa,0)}^{-1}\hat{\bf x} )^\dagger
({\bf C}+{\bf N})^{-1}
\Lambda_{(\kappa,0)}^{-1}\hat{\bf x}
+ \kappa^\dagger{\bf C}^{\kappa\kappa\,-1}\kappa.
\label{eq:chisq}
\end{equation}
Here $\Lambda_{(\kappa,0)}^{-1}\hat{\bf x}$ is the de-lensed CMB, i.e. we
have taken the measured CMB and projected it back onto the primary CMB
assuming that the lens configuration is given by convergence $\kappa$ with
no rotation.  To a first approximation, this should have covariance ${\bf
C}+{\bf N}$ since it is the sum of primary CMB and instrument noise.  
(The matrix ${\bf C}+{\bf N}$ is equal to the measured CMB covariance in
the absence of lensing and hence will frequently be denoted by $\hat{\bf
C}_{(0,0)}$.  Technically the noise covariance is not exactly ${\bf N}$
because the noise has been de-lensed; see \S\ref{sec:estcon}.) 
We have thus chosen to define our
$\chi^2$ as the amount of power in this de-lensed CMB, with the various
modes weighted according to their variance.  The addition of the
$\kappa^\dagger{\bf C}^{\kappa\kappa\,-1}\kappa$ term serves to regularize
the problem by preventing the convergence from running off to $\infty$ in
search of smaller primary CMB power.

If we take the first-order approximation to $\Lambda^{-1}$ given by Eq.
(\ref{eq:lambdaapprox}), Eq. (\ref{eq:chisq}) becomes:
\begin{equation}
\chi^2(\kappa) = \chi^2(0) + 2 \sum_\bfl m_\bfl^\ast \kappa_\bfl 
+ \sum_{\bfl,\bfl'}\kappa_\bfl^\ast (A_{\bfl,\bfl'} +
C^{\kappa\kappa\,-1}_l\delta_{\bfl,\bfl'})\kappa_{\bfl'},
\label{eq:chisq2}
\end{equation}
where:
\begin{eqnarray}
\chi^2(0) = && \hat{\bf x}^\dagger({\bf C}+{\bf N})^{-1} \hat{\bf x}, 
\nonumber \\
m_\bfl = && 
\hat{\bf x}^\dagger({\bf C}+{\bf N})^{-1} \sigma^\kappa_{-\bfl} \hat{\bf x},
\nonumber \\
A_{\bfl,\bfl'} = && 
\hat{\bf x}^\dagger \sigma^\kappa_{-\bfl} ({\bf C}+{\bf N})^{-1} 
\sigma^\kappa_{\bfl'} \hat{\bf x}.
\label{eq:chisq2b}
\end{eqnarray}
Note that ${\bf m}$ is a real vector and ${\bf A}$ is Hermitian. This is a
quadratic function of $\kappa$ and hence it has a minimum that can be
determined via standard techniques.  The minimum is at:
\begin{equation}
\kappa_\ast = ({\bf A}+{\bf C}^{\kappa\kappa\,-1})^{-1}{\bf m}.
\label{eq:k1}
\end{equation}
The error covariance ${\bf S}_{\chi^2}$ of $\kappa$ is found by the usual
method of setting
$\chi^2(\kappa)=\chi^2(\kappa_\ast)+(\kappa-\kappa_\ast)^\dagger{\bf
S}_{\chi^2}^{-1}(\kappa-\kappa_\ast)$; this yields:
\begin{equation}
{\bf S}_{\chi^2} = ({\bf A}+{\bf C}^{\kappa\kappa\,-1})^{-1}.
\label{eq:k2}
\end{equation}

The most important feature of this analysis is the reconstruction error,
${\bf S}_{\chi^2}$.  Note that as the instrument noise goes to zero, the
matrix ${\bf C}+{\bf N}$ develops null directions corresponding to the
$B$-modes.  Therefore, $({\bf C}+{\bf N})^{-1}$ has infinite eigenvalues
in these directions, and if the number of convergence modes being
reconstructed is less than or equal to the number of $B$-modes measured,
we have ${\bf A}\rightarrow\infty$ and ${\bf S}_{\chi^2}\rightarrow 0$.  
This leads us to the conclusion that the accuracy of convergence
reconstruction is limited only by the sensitivity of the instrument and
the presence of foregrounds or other contaminants, not by statistics of
the convergence or primary CMB field.  One can note that for zero
instrument noise, the $\chi^2$, Eq. (\ref{eq:chisq}), is infinite unless
the de-lensed CMB field $\Lambda_{(\kappa,0)}^{-1}\hat{\bf x}$ has
vanishing $B$-modes, i.e. in this case the $\chi^2$ analysis is solving
for $B_{\rm unlensed}=0$ (except possibly for a few degenerate modes; see
Appendix \ref{sec:bmode}).  We extend this methodology to a full
likelihood analysis in \S\ref{sec:likelihood}, where we find that the
general conclusions of this section remain valid.

\section{Lensing reconstruction: likelihood analysis}
\label{sec:likelihood}

In this section we explore the accuracy of reconstruction of lensing based
on CMB temperature and polarization.  We follow the analysis performed in
Ref.  \cite{2003PhRvD..67d3001H} for the CMB temperature; most of the
analysis extends easily to polarization, with one exception: the primary
CMB has very little (if any) $B$-mode polarization.  This means that the
lensed CMB power spectrum $\langle \tilde {\bf C}^{BB}_l \rangle_{LSS}$
cannot be expressed as a small perturbation on the unlensed power
spectrum.  We also include the effect of the field rotation in our
discussion of the likelihood gradient and Fisher matrix, although we do
not construct a ``practical'' estimator for it.

\subsection{Likelihood function and gradient}
\label{sec:gradient}

For a given lens configuration with re-mapping function $g$, the
covariance matrix $\hat {\bf C}$ of the measured CMB is computed from:
\begin{equation}
\hat{\bf C}_g = \langle \hat{\bf x}\hat{\bf x}^\dagger \rangle
= \langle (\tilde{\bf x}+\eta)(\tilde{\bf x}+\eta)^\dagger \rangle =
\tilde{\bf C}_g+{\bf N} = \Lambda_g {\bf C} \Lambda_g^\dagger + {\bf N},
\label{eq:chat}
\end{equation}
where ${\bf N}=\langle \eta\eta^\dagger\rangle$ is the noise matrix.  The
measured CMB is Gaussian-distributed if we assume that the primary CMB
${\bf x}$ and instrument noise $\eta$ are both Gaussian.  (Note: the
assumption of Gaussianity only applies {\em before} we average over LSS
realizations.) The (negative log) likelihood function ${\cal L}$ for a
lens configuration with re-mapping function $g$ is then given (up to an
irrelevant constant) by:
\begin{equation}
{\cal L}(g) = {1\over 2}\ln\det\hat{\bf C}_g 
+ {1\over 2}\hat{\bf x}^\dagger\hat{\bf C}^{-1}_g \hat{\bf x},
\label{eq:likelihood}
\end{equation}
Now we wish to determine the likelihood gradient with respect to the lens
configuration $g=(\kappa,\omega)$.  We will compute the gradient of the
likelihood function, Eq. (\ref{eq:likelihood}), using Eq. (\ref{eq:chat}):
\begin{equation}
{\partial{\cal L}\over \partial\kappa_\bfl} = \Tr\left( \hat{\bf C}^{-1}_g
{\partial\Lambda_g\over \partial\kappa_\bfl[g]}
{\bf C} \Lambda_g^\dagger \right)
- \hat{\bf x}^\dagger \hat{\bf C}^{-1}_g
{\partial\Lambda_g\over \partial\kappa_\bfl[g]} {\bf C}\Lambda_g^\dagger
\hat{\bf C}^{-1}_g \hat{\bf x}.
\label{eq:lgrad}
\end{equation}

The maximum-likelihood estimator is given by the relation $\partial{\cal
L}/\partial\kappa_\bfl = 0$.  (We also require $\partial{\cal
L}/\partial\omega_\bfl = 0$ if we are estimating $\omega$ as well as
$\kappa$.)  However, maximum likelihood estimation of the lensing field is
generally unstable because the lensing field has too many degrees of
freedom.  In order to regularize the problem, we introduce a Bayesian
prior probability distribution $\propto e^{-\wp}$ for $g=(\kappa,\omega)$,
i.e. we take prior probability $dP \propto e^{-\wp(g)} \prod_\bfl
d\kappa_\bfl d\omega_\bfl$. It is most convenient to take a Gaussian prior
based on the power spectra of $\kappa$ and (if applicable) $\omega$:

\begin{eqnarray}
\wp(\kappa,\omega) = &&
{1\over 2}\left( \kappa^\dagger {\bf C}^{\kappa\kappa\,-1}\kappa 
+ \ln\det{\bf C}^{\kappa\kappa}\right)
+ {1\over 2}\left( \omega^\dagger {\bf C}^{\omega\omega\,-1}\omega 
+ \ln\det{\bf C}^{\omega\omega}\right)
\nonumber \\ = &&
{1\over 2}\sum_\bfl \left( {|\kappa_\bfl|^2\over C^{\kappa\kappa}_l} 
+ \ln C^{\kappa\kappa}_l \right) +
{1\over 2}\sum_\bfl \left( {|\omega_\bfl|^2\over C^{\omega\omega}_l} 
+ \ln C^{\omega\omega}_l \right),
\end{eqnarray}
where in the second equality we have assumed that the prior on $\kappa$
and $\omega$ is statistically isotropic. (Note that this assumes the power
spectra are known; we will consider the problem of estimating
$C^{\kappa\kappa}_l$ from CMB data in \S\ref{sec:ape}.  The methods we
present in \S\ref{sec:ape} allow iterative determination of both the
convergence field $\kappa$ and the power spectrum $C^{\kappa\kappa}_l$.)
If we are neglecting the field rotation then the terms involving $\omega$
should
simply be removed. The mode of the posterior probability distribution is
given by minimizing ${\cal L}+\wp$; we thus set
$\partial\wp/\partial\kappa_\bfl = -\partial{\cal L}/\partial\kappa_\bfl$,
or:
\begin{equation}
[{\bf C}^{\kappa\kappa\,-1}\kappa]_\bfl^\ast =
-\Tr\left( \hat{\bf C}^{-1}_g
{\partial\Lambda_g\over \partial\kappa_\bfl[g]}
{\bf C} \Lambda_g^\dagger \right)
+ \hat{\bf x}^\dagger \hat{\bf C}^{-1}_g
{\partial\Lambda_g\over \partial\kappa_\bfl[g]}
{\bf C} \Lambda_g^\dagger \hat{\bf C}^{-1}_g \hat{\bf x}.
\label{eq:wf}
\end{equation}
Because of the presence of the prior ${\bf C}^{\kappa\kappa\,-1}$, this
estimator will filter out lensing modes that cannot be accurately
reconstructed from the CMB data.  It can thus be viewed as a sort of
nonlinear generalization of the Wiener filter.

\subsection{Practical estimator for the convergence}
\label{sec:estcon}

The likelihood gradient, Eq. (\ref{eq:lgrad}), and hence the convergence
estimator Eq. (\ref{eq:wf}) based on it, are difficult to evaluate.  We
therefore investigate several approximations to the likelihood function.  
First, we consider only the convergence, $\kappa$; the rotation $\omega$
will be shown in \S\ref{sec:crosspotential} to be unimportant unless
instrument noise is very small.  We note that Eq. (\ref{eq:wf}) can be
re-written as:
\begin{equation}
[{\bf C}^{\kappa\kappa,-1}\kappa]_\bfl^\ast
= -\Tr\left( \Lambda_g^{\dagger\,-1}{\bf w} \Lambda_g^{-1}
{\partial\Lambda_g\over \partial\kappa_\bfl[g]} {\bf C} 
{\bf w}\Lambda_g^{-1} \hat{\bf C}_g \right)
+ \hat{\bf x}^\dagger \Lambda_g^{\dagger\,-1}{\bf w} \Lambda_g^{-1}
{\partial\Lambda_g\over \partial\kappa_\bfl[g]}{\bf C} {\bf w}\Lambda_g^{-1}
\hat{\bf x},
\label{eq:wf2}
\end{equation}
where the weight matrix ${\bf w}$ is defined by:
\begin{equation}
{\bf w} = \Lambda_g^\dagger\hat{\bf C}_g^{-1}\Lambda_g 
= ({\bf C} + \Lambda_g^{-1}{\bf N}\Lambda_g^{\dagger\,-1})^{-1}.
\label{eq:ape2b}
\end{equation}
Here $\Lambda_g^{-1}{\bf N}\Lambda_g^{\dagger\,-1}$ is the de-lensed noise
covariance matrix, which is equal to the noise covariance ${\bf N}$ for
$g=0$ (no de-lensing).  Under most circumstances, de-lensing has
much less effect on the noise than on the CMB signal, because instrument
noise is a relatively smooth function of $l$ and contains both $E$
and $B$ modes with similar power.  It is possible that $\Lambda_g^{-1}{\bf
N}\Lambda_g^{\dagger\,-1}\not\approx{\bf N}$ if the noise power spectrum
contains sharp features; in this case, the approximation ${\bf w}\approx
\hat{\bf C}_{(0,0)}^{-1}$ used below
may result in a non-optimal, or (in extreme cases) unstable estimator.

We would like to approximate $\Lambda_g^{-1}(\partial\Lambda_g /
\partial\kappa_\bfl[g])$ using the $\sigma$ matrices; this can be done by
expanding:
\begin{equation}
\Lambda_g^{-1} {\partial\Lambda_g\over \partial\kappa_\bfl[g]}
= \left. {\partial \kappa_{\bfl'}[g^{-1}g']\over \partial \kappa_\bfl[g']}
\right|_{g'=g} \sigma^\kappa_\bfl
+ \left. {\partial \omega_{\bfl'}[g^{-1}g']\over \partial
\kappa_\bfl[g']}\right|_{g'=g} \sigma^\omega_\bfl,
\end{equation}
where the juxtaposition $g^{-1}g'$ indicates composition of the lensing
operations: $(g^{-1}g')X = g^{-1}(g'(X))$. If the lensing is very weak we
may take $(\partial
\kappa_{\bfl'}[g^{-1}g']/\kappa_\bfl[g'])|_{g'=g}\approx
[\Lambda_g^{-1}]_{\bfl',\bfl}$ and $(\partial
\omega_{\bfl'}[g^{-1}g']/\kappa_\bfl[g'])|_{g'=g}\approx 0$, that is, the
composition of lensing operations can be approximated by re-mapping the
convergence field and neglecting rotation. In this regime, the statistical
properties of the convergence field $\kappa$ should not differ greatly
from those of the ``de-lensed'' convergence field $\Lambda_g^{-1}\kappa$;
mathematically, this means that we may take $\Lambda_g$ and ${\bf
C}^{\kappa\kappa}$ to commute. With these approximations, Eq.
(\ref{eq:wf2}) becomes:
\begin{equation}
C_l^{\kappa\kappa\,-1} (\Lambda_g\kappa)_\bfl^\ast
= (\Lambda_g^{-1}\hat{\bf x})^\dagger {\bf w} \sigma^\kappa_\bfl {\bf C}
{\bf w} \Lambda_g^{-1} \hat{\bf x}
-\Tr\left[ {\bf w} \sigma^\kappa_\bfl {\bf C} {\bf w} \Lambda_g^{-1}
\hat{\bf C}_g \Lambda_g^{\dagger\,-1} \right].
\label{eq:ape2}
\end{equation}

The right-hand side of Eq. (\ref{eq:ape2}) is our approximation to the
likelihood gradient, and the left-hand side is our (approximate) prior
gradient.  Note that the right-hand side evaluated at the correct lensing
configuration $g$ has expectation value zero, regardless of the choice of
weight function; we will therefore choose the slightly sub-optimal weight
function ${\bf w} = \hat{\bf C}_{(0,0)}^{-1}$ in order to reduce
computational difficulties. This leads us to the estimator:
\begin{equation}
C_l^{\kappa\kappa\,-1} (\Lambda_g\kappa)_l^\ast
= (\hat{\bf C}_{(0,0)}^{-1} \Lambda_g^{-1}\hat{\bf x})^\dagger
\sigma^\kappa_\bfl {\bf C} \hat{\bf C}_{(0,0)}^{-1} 
\Lambda_g^{-1}\hat{\bf x}
- \Tr\left[
\Lambda_g^{\dagger\,-1}\hat{\bf C}_{(0,0)}^{-1} \sigma^\kappa_\bfl {\bf C}
\hat{\bf C}_{(0,0)}^{-1} \Lambda_g^{-1} \hat{\bf C}_g \right].
\label{eq:estcon0}
\end{equation}
[This choice leads to some difficulty for low-noise, wide-beam
($\theta_{FWHM}\ge 10$ arcmin) experiments; see \S\ref{sec:numer1} for
details.] By expanding $\hat{\bf C}_g$ using Eq. (\ref{eq:chat}), and
noting that in the harmonic-space basis, ${\bf C}$ and $\hat{\bf
C}_{(0,0)}$ are diagonal whereas $\sigma^\Phi_\bfl$ has no nonzero
diagonal elements, we convert this into:
\begin{equation}
C_l^{\kappa\kappa\,-1} (\Lambda_g\kappa)_\bfl^\ast
= (\hat{\bf C}_{(0,0)}^{-1} \Lambda_g^{-1}\hat{\bf x})^\dagger
\sigma^\kappa_\bfl {\bf C}\hat{\bf C}_{(0,0)}^{-1} 
\Lambda_g^{-1}\hat{\bf x}
- \Tr\left[
\Lambda_g^{\dagger\,-1}\hat{\bf C}_{(0,0)}^{-1}
\sigma^\kappa_\bfl {\bf C} \hat{\bf C}_{(0,0)}^{-1} \Lambda_g^{-1}
{\bf N} \right].
\label{eq:estcon}
\end{equation}

\subsection{Fisher matrix}

The Fisher matrix is defined as the expectation value of the second
derivative of the likelihood function:
\begin{equation}
F[\kappa_\bfl,\kappa_{\bfl'}] \equiv
\left.\left\langle {\partial^2{\cal L}\over 
\partial\kappa_\bfl^\ast \partial\kappa_{\bfl'}}
\right\rangle\right|_{(\kappa,\omega)}
= \left.\left\langle {\partial{\cal L}\over\partial\kappa_\bfl^\ast} 
{\partial{\cal L}\over\partial\kappa_{\bfl'}}
\right\rangle\right|_{(\kappa,\omega)},
\label{eq:fisherdef}
\end{equation}
where the second equality follows from taking the second derivative
($\partial^2/\partial\kappa_\bfl^\ast\partial\kappa_{\bfl'}$) of the
normalization condition $\int e^{-\cal L}{\cal D}\hat{\bf x} = 1$ and
noting that the expectation value of any statistic ${\cal S}$ is
$\langle{\cal S}\rangle = \int {\cal S} e^{-\cal L}{\cal D}\hat{\bf x}$.  
(This also shows that ${\bf F}$ has all non-negative eigenvalues.) A
similar relation holds for the field-rotation modes $\omega_\bfl$. We may
thus compute the lensing Fisher matrix as the covariance of the likelihood
gradient; the easiest method of doing this is to apply Wick's theorem to
compute the variance of Eq. (\ref{eq:lgrad}). This yields:
\begin{equation}
F[\kappa_\bfl,\kappa_{\bfl'}] = 
\left\langle {\partial{\cal L}\over\partial\kappa_\bfl^\ast} {\partial{\cal 
L}\over\partial\kappa_{\bfl'}} \right\rangle
= \Tr\left( \Lambda_g^\dagger \hat{\bf C}^{-1}_g
\sigma^\kappa_{-\bfl} {\bf C} \Lambda_g^\dagger \hat{\bf C}^{-1}_g
\sigma^\kappa_{\bfl'} {\bf C} \right)
+ \Tr\left( \hat{\bf C}^{-1}_g
\sigma^\kappa_{-\bfl} {\bf C} \Lambda_g^\dagger
\hat{\bf C}^{-1}_g \Lambda_g {\bf C} \sigma^{\kappa\dagger}_{\bfl'}
\right)
\label{eq:fisher}
\end{equation}
and similarly for the components of the Fisher matrix elements involving
the field rotation.  For simplicity, we compute the Fisher matrix at
$g=(0,0)$, i.e. the $\kappa=\omega=0$ point, so that $\Lambda_{(0,0)}$ is
the identity. The ${\bf C}$ and $\hat{\bf C}_{(0,0)}^{-1}$ matrices are
diagonal in the $\{T,E,B\}$ basis:
\begin{equation}
{\bf C} = \left( \begin{array}{ccc} C_l^{TT} & C_l^{TE} & 0 \\
C_l^{TE} & C_l^{EE} & 0 \\
0 & 0 & C_l^{BB} \end{array} \right),
\end{equation}
and:
\begin{equation}
\hat{\bf C}^{-1}_{(0,0)} =
\left( \begin{array}{ccc}
(C_l^{EE}+N_l^{EE})/D_l & -C_l^{TE}/D_l & 0 \\
-C_l^{TE}/D_l & (C_l^{TT}+N_l^{TT})/D_l & 0 \\
0 & 0 & 1/(C_l^{BB}+N_l^{BB})
\end{array} \right),
\label{eq:u}
\end{equation}
where $D_l = (C_l^{TT}+N_l^{TT})(C_l^{EE}+N_l^{EE})-(C_l^{TE})^2$.  

The overall Fisher matrix is then computed from Eq. (\ref{eq:fisher}):
\begin{equation}
F[\kappa_\bfl,\kappa_{\bfl'}] \approx
\frac{1}{2} \Tr \left[ \hat{\bf C}^{-1}_{(0,0)} 
{\bf f}_{-\bfl}^\kappa \hat{\bf C}^{-1}_{(0,0)}
{\bf f}_{\bfl'}^\kappa \right],
\label{eq:fphi}
\end{equation}
where:
\begin{eqnarray}
\!\!
[{\bf f}^\kappa_\bfl]_{\bfl_1,-\bfl_2} = && \!\!\!\!
[{\bf C}\sigma_{-\bfl}^{\kappa\,\dagger} 
+ \sigma_\bfl^\kappa{\bf C}]_{\bfl_1,\bfl_2}
\nonumber\\
= && \!\!\!\! -{\delta_{\bfl_1,\bfl-\bfl_2}\over\sqrt{4\pi}} 
\left( \frac{2}{l^2}\right) \bfl\cdot
\left( \begin{array}{ccc}
\bfl_1 C^{TT}_{l_1} + \bfl_2 C^{TT}_{l_2} &
\bfl_1 C^{TE}_{l_1} \cos 2\alpha + \bfl_2 C^{TE}_{l_2}
& -\bfl_1 C^{TE}_{l_1} \sin 2\alpha \\
\bfl_1 C^{TE}_{l_1} +
\bfl_2 C^{TE}_{l_2} \cos 2\alpha & 
(\bfl_1 C^{EE}_{l_1}+\bfl_2 C^{EE}_{l_2}) \cos 2\alpha &
( \bfl_2 C^{BB}_{l_2} - \bfl_1 C^{EE}_{l_1}) \sin 2\alpha \\
\bfl_2 C^{TE}_{l_2} \sin 2\alpha 
& (\bfl_2 C^{EE}_{l_2}-\bfl_1 C^{BB}_{l_1}) \sin 2\alpha 
& (\bfl_1 C^{BB}_{l_1}+\bfl_2 C^{BB}_{l_2}) \cos 2\alpha
\end{array} \right)
.
\label{eq:fl}
\end{eqnarray}
(The matrix ${\bf f}^\omega_\bfl$ is identical except for the replacement
$\bfl\cdot\rightarrow\star\bfl\cdot$.) Note that by Hermiticity of ${\bf
C}$, we have ${\bf f}^\kappa_\bfl = {\bf f}^{\kappa\,\dagger}_{-\bfl}$;
for the individual $3\times 3$ blocks in the harmonic-space basis, $[{\bf
f}^\kappa_\bfl]_{\bfl_1,-\bfl_2} = [{\bf
f}^{\kappa\,\dagger}_{-\bfl}]_{-\bfl_1,\bfl_2}^\dagger$. Also our
construction guarantees that ${\bf f}^\kappa_\bfl = \partial\hat{\bf
C}/\partial\kappa_\bfl$ where the derivative is evaluated at
$\kappa=\omega=0$.

It can be verified by explicit matrix multiplication that the computation
for $F[\kappa,\kappa]$ here yields the uncertainty in the minimum-variance
quadratic estimator of Ref. \cite{2002ApJ...574..566H}, with one
exception: we have computed the Fisher matrix at $g=0$, hence the
denominator of Eq. (\ref{eq:fphi}) contains the {\em unlensed} CMB power
spectrum plus the instrument noise, whereas the equivalent calculation in
Ref. \cite{2002ApJ...574..566H} contains the {\em lensed} CMB power
spectrum plus the instrument noise.  In the case of quadratic estimation,
it is clear that the lensed power spectrum should be used in order to
minimize the variance of the estimator.  Conceptually, this is because the
lensing $B$-modes can be iteratively cleaned from the map, thereby
reducing the post-cleaning $B$-mode power spectrum and reducing the noise
in the lensing estimator.  Our ability to clean the map is bounded, of
course, by the sum of the unlensed CMB and noise contributions to
$C^{BB}_l$.

\subsection{Uncertainty in lens reconstruction}
\label{sec:deg}

The usual method of estimating the uncertainty in lens reconstruction
would be to invert the Fisher matrix.  This approach is motivated by the
Cramer-Rao inequality, which states that an unbiased estimator of the
lensing configuration must have covariance at least equal to ${\bf
F}^{-1}$.  Unfortunately, the Cramer-Rao inequality is only an inequality,
and there is no guarantee that the bound ${\bf F}^{-1}$ can actually be
reached; indeed this bound is only achieved in the case where the
likelihood function is Gaussian with curvature ${\bf F}$.  The traditional
justification for assuming Gaussianity of the likelihood function is the
Central Limit Theorem.  This works for studies of lensing of the CMB
temperature field, in which the typical lensing mode being reconstructed
is at $l\sim 100$ whereas the temperature fluctuations that are being
lensed have wavenumber $l\sim 1000$; thus there are roughly
$(1000/100)^2=100$ patches of primary CMB behind each lensing mode.  
(Most of the information comes from non-local correlations in the $T$
field, so this argument technically requires more justification;
nevertheless the calculations in Ref. \cite{2003PhRvD..67d3001H} seem to
indicate that it gives the correct answer.)  This argument does not apply
to lensing of the CMB polarization because the wavenumbers of the primary
$E$ polarization modes and of the lensing field modes ($\kappa_l$) are
both at wavenumbers of order $l\sim 1000$.  We should therefore be careful
of possible problems with the Fisher matrix estimate, Eq. (\ref{eq:fphi})
of the uncertainty in the lensing field.  In this section, we outline two
such problems that occur in lensing reconstruction: first, a complete
breakdown of the Fisher matrix approach when the field rotation $\omega$
becomes important; and second, fluctuations in the curvature matrix
resulting from the statistical nature of the primary $E$-field.

Consider first the problem of simultaneous reconstruction of both $\kappa$
and $\omega$.  (We will see in \S\ref{sec:crosspotential} that the noise
levels required for this are not achievable in the near term, however,
this extreme example serves to illustrate the problem.)  One can see that
if there are no primary $B$-modes, then as instrument noise goes to zero,
the uncertainty in $\kappa$ and $\omega$ obtained by inverting Eq.
(\ref{eq:fphi}) goes to zero.  But this cannot be true because the {\em
one} equation $B_{\rm unlensed}=0$ cannot be used to solve for the {\em
two} fields $\kappa$ and $\omega$ simultaneously.  Therefore inverting the
Fisher matrix yields a qualitatively absurd conclusion. What went wrong?  
The observation that one equation (the vanishing of the unlensed $B$-mode
field) cannot be solved for two variables ($\kappa$ and $\omega$) yields a
clue.  Consider the case where instrument noise is negligible; then we
know that the measured $B$-mode is purely caused by lensing:
\begin{equation}
\hat B_\bfl =\sum_{\bfl'} {1\over\sqrt{4\pi}} \left( \frac{2}{{l'}^2}\right)
\left[\bfl'\cdot(\bfl-\bfl') \kappa_{\bfl'} 
+ \star\bfl'\cdot(\bfl-\bfl') \omega_{\bfl'}\right]
E_{\bfl-\bfl'}\sin 2\alpha,
\label{eq:dege}
\end{equation}
where $\alpha = \phi_\bfl - \phi_{\bfl-\bfl'}$. To the extent that the
$E$-mode is dominated by the primary (not lensing) contribution, Eq.
(\ref{eq:dege}) is a linear system containing $2N$ unknown variables (the
amplitudes of the $\kappa$ and $\omega$ modes) but only $N$ equations (the
knowledge of the lensed $B$-modes), thus there are degeneracy directions
in lens configuration space which are unconstrained by the vanishing of
the $B$-modes.  These directions must be constrained by a combination of
the statistical properties of primary temperature and $E$-type
polarization, and prior knowledge about the lensing field.

We can now understand why the lensing Fisher matrix ${\bf F}$ is
inadequate for determining the uncertainty in the lensing fields $\kappa$
and $\omega$.  The curvature matrix:
\begin{equation}
{\cal F}[\kappa_\bfl,\kappa_{\bfl'}](g) 
= {\partial^2{\cal L}\over \partial \kappa_\bfl^\ast \partial\kappa_{\bfl'}}
\label{eq:curvaturematrix}
\end{equation}
has very small eigenvalues in the directions of degeneracy of Eq.
(\ref{eq:dege}) and very large eigenvalues (approaching $\infty$ as ${\bf
N}\rightarrow 0$) in the orthogonal directions.  But as one can see from
Eq.  (\ref{eq:dege}), the direction of degeneracy depends on $E$ and hence
on the specific realization of the CMB.  If we average over CMB
realizations to obtain a Fisher matrix ${\bf F}$, then we derive ${\bf
F}=\infty$, which does not accurately reflect the non-zero errors in the
degenerate directions of Eq. (\ref{eq:dege}).  Mathematically, the Fisher
matrix methodology does not work because the error bars on
$(\kappa,\omega)$ are extremely non-Gaussian.  The lesson is that we
should be careful about interpreting the inverse of the Fisher matrix as
an uncertainty in parameters when the Central Limit Theorem does not come
to our aid.

A similar but less spectacular problem occurs in attempting reconstruction
of small-scale lensing modes even when there is sufficient instrument
noise that $\Omega$ is irrelevant. This is the regime of interest to a
near-future high-resolution polarization experiment. The statistical
uncertainty in the lensing reconstruction is given by the inverse of the
curvature matrix ${\cal F}$.  When doing a lens reconstruction, this
curvature matrix is augmented by the curvature of the prior, ${\bf
C}^{\kappa\kappa\,-1}$, so that the posterior error covariance matrix of
the lensing reconstruction is approximately $({\cal F}+{\bf
C}^{\kappa\kappa\,-1})^{-1}$.  We wish to compute the mean squared error
in the reconstructed convergence $\hat\kappa$, which is obtained by
computing the ensemble average of this covariance matrix over all
realizations of CMB, noise, and LSS:
\begin{equation}
{\bf S}^{\kappa\kappa} 
= \langle (\hat\kappa - \kappa)(\hat\kappa-\kappa)^\dagger \rangle_{LSS}
\approx\langle ({\cal F}+{\bf C}^{\kappa\kappa\,-1})^{-1} \rangle_{LSS}.
\label{eq:sdef}
\end{equation}
The Fisher matrix ${\bf F}$ is defined to be the expectation value of the
curvature: ${\bf F}=\langle{\cal F}\rangle$ (with no LSS average). If the
curvature matrix were always equal to ${\bf F}$, then it would be
permissible to approximate ${\bf S}^{\kappa\kappa} \approx ({\bf F}+{\bf
C}^{\kappa\kappa\,-1})^{-1}$.  It can be shown (see Appendix \ref{sec:a1})
that the statistical fluctuations of ${\cal F}$ always increase the
uncertainty, Eq. (\ref{eq:sdef}); this increase we call the ``curvature
correction.''

Conceptually, the naive calculation that the mean squared error is
approximately ${\bf S}_0=({\bf F}+{\bf C}^{\kappa\kappa\,-1})^{-1}$
suffers problems for the same reason that the Fisher matrix calculation
for simultaneously estimating $\kappa$ and $\omega$ failed: the different
realizations of the primary CMB introduce fluctuations in ${\cal F}$, and
when we average over CMB realizations we generate a non-Gaussian error
distribution for the estimated convergence.

The actual computation of the curvature corrections is not pursued here;
some of the relevant issues are discussed in Appendix \ref{sec:a1}, where
we show that the ``first-order noise contribution'' of Ref.
\cite{2003astro.ph..2536K} arises as one part of the second-order
curvature correction.

\subsection{Relation to quadratic estimators}
\label{sec:quad}

It is of interest to compare the estimator we have derived, Eq.
(\ref{eq:estcon}), to the quadratic estimation method of Ref.
\cite{2002ApJ...574..566H}.  The performance of the estimators is compared
numerically in \S\ref{sec:estcon}.  Here we display the quadratic
estimator and note the major differences between the quadratic and
iterative estimators.  The Wiener-filtered quadratic estimator is:
\begin{equation}
\hat \kappa_\bfl^\ast = 
\left[
C_l^{\kappa\kappa\,-1} \delta_{\bfl,\bfl'} + F^{\rm (quad)}_{\bfl,\bfl'}
\right]^{-1}
\hat{\bf x}^\dagger \langle\hat{\bf C}\rangle_{LSS}^{-1} 
\sigma^\kappa_{\bfl'} {\bf C}  \langle\hat{\bf C}\rangle_{LSS}^{-1}
\hat{\bf x},
\label{eq:wfquad}
\end{equation}
where the quadratic Fisher matrix is determined as:
\begin{equation}
F^{\rm (quad)}_{\bfl,\bfl'} 
= {1\over 2}\Tr\left[
\langle\hat{\bf C}\rangle_{LSS}^{-1} {\bf f}_{-\bfl}^\kappa
\langle\hat{\bf C}\rangle_{LSS}^{-1} {\bf f}_{\bfl'}^\kappa
\right].
\label{eq:qfisher}
\end{equation}
The ``unbiased'' (to first order in $\Phi$), non-Wiener-filtered
temperature is given by Eq. (\ref{eq:wfquad}) with the ``prior term''
$C_l^{\kappa\kappa\,-1} \delta_{0,\bfl+\bfl'}$ omitted.  (The quadratic
Fisher matrix is not technically a Fisher matrix, but its inverse does
give the covariance of the unbiased quadratic estimator.) We prove in
Appendix \ref{sec:hu} that Eq. (\ref{eq:wfquad}) and its unbiased
equivalent are identical to the minimum-variance quadratic estimator that
arises from the optimal weighting scheme of Ref.
\cite{2002ApJ...574..566H}.  The mean squared error in the reconstructed
convergence according to Eq. (\ref{eq:wfquad}) is:
\begin{equation}
{\bf S}^{\kappa\kappa\,\rm (quad)} 
= \left( {\bf C}^{\kappa\kappa\,-1} + {\bf F}^{\rm (quad)} \right)^{-1}.
\label{eq:squad}
\end{equation}

Several features of Eqs. (\ref{eq:wfquad}) through (\ref{eq:squad}) are
readily apparent.  First, the estimator Eq. (\ref{eq:wfquad}) is a
quadratic function of the CMB temperature/polarization field ${\bf x}$.
Secondly, we note that the uncertainty in the quadratic estimator is
determined by the quadratic Fisher matrix, which contains the inverse of
$\langle\hat{\bf C}\rangle_{LSS}$.  For statistically isotropic noise,
this inverse is given by:
\begin{equation}
\langle\hat{\bf C}\rangle_{LSS}^{-1} =
\left( \begin{array}{ccc}
(\tilde C_l^{EE}+N_l^{EE})/\tilde D_l & -\tilde C_l^{TE}/\tilde D_l & 0 \\
-\tilde C_l^{TE}/\tilde D_l & (\tilde C_l^{TT}+N_l^{TT})/\tilde D_l & 0 \\
0 & 0 & 1/(\tilde C_l^{BB}+N_l^{BB})
\end{array} \right),
\label{eq:uquad}
\end{equation}
where $\tilde D_l = \tilde C_l^{TT}\tilde C_l^{EE}-\tilde C_l^{TE\,2}$,
and $\tilde C_l^{XX'}$ is the lensed CMB power spectrum (or
cross-spectrum): $\tilde C_l^{XX'} = \langle \tilde X_\bfl \tilde
{X'}_\bfl^{\ast} \rangle_{LSS}$.  Comparison of the quadratic Fisher
matrix [Eq. (\ref{eq:qfisher})] to the full Fisher matrix [Eq.
(\ref{eq:fphi})] shows that the two are identical except for replacement
of Eq. (\ref{eq:uquad}) by Eq. (\ref{eq:u}).  This results in a
qualitative difference between the two estimators: as instrument noise is
reduced toward zero, the full Fisher matrix improves without bound (${\bf
F}\rightarrow \infty$), so (aside from foregrounds, field rotation,
primary $B$-modes, and the statistical concerns outlined in
\S\ref{sec:deg}) the iterative estimator should be able to reconstruct the
convergence with arbitrary accuracy.  This is not so for the quadratic
estimator, whose reconstruction accuracy is limited by the nonzero value
of $\tilde C_l^{BB}$ and the resulting upper bounds on $\langle\hat{\bf
C}\rangle_{LSS}^{-1}$ and ${\bf F}^{\rm (quad)}$.  At high noise levels
where the $B$-mode cannot be mapped, however, $C_l^{BB}+N_l^{BB}\approx
\tilde C_l^{BB}+N_l^{BB}$ since both sides of the equation are
noise-dominated, and in this regime the performance of the two estimators
should be nearly identical.

\section{Statistics of field rotation}
\label{sec:crosspotential}

Here we investigate the statistics of weak lensing fields with the
objective of understanding the importance of the field rotation $\omega$
(or equivalently the cross-potential $\Omega$) in CMB weak lensing.  
Field rotation is a cosmological contaminant in the sense that even with
noiseless CMB data and no foregrounds, we cannot hope to recover two
fields $\kappa$ and $\omega$ from the single equation $B_{\rm
unlensed}=0$.  Therefore a non-zero power spectrum $C^{\omega\omega}_l$
translates into an uncertainty in the lens reconstruction. We compute the
power spectrum $C^{\omega\omega}_l$ by considering deflection angles; this
has the advantage of providing a unified treatment of the higher-order
Born approximation and ``lens-lens coupling'' effects considered by Ref.
\cite{2002ApJ...574...19C}. We work in the longitudinal gauge because in
this gauge the perturbations to the metric remain small (of order
$10^{-5}$ except in the very small portion of the universe near neutron
stars and black holes) and so perturbation theory techniques are valid. We
then consider the implications for lensing reconstruction; for near-term
experiments, the effect is seen to be negligible.

\subsection{Lensing power spectra}
\label{sec:scalarlens}

In the flat-sky approximation, we treat the photons as propagating in
roughly the $-\hat{\bf e}_z$ direction so that the CMB experiment looks in
the $\hat{\bf e}_z$ direction; the ``sky'' is in the $xy$-plane.  The
spacetime metric observed by the photon is (so long as it does not stray
far from the $z$-axis):
\begin{equation}
ds^2 = a^2(\tau) \left[ -(1+2\Psi)d\tau^2 +
(1-2\Psi)(d\chi^2+\sin_K^2\chi\; 
(d\hat n_x^2+d\hat n_y^2)) \right] ,
\end{equation}
where the Newtonian potential $\Psi$ is generated by the non-relativistic
matter inhomogeneities, and $\sin_K\chi=K^{-1/2}\sin(K^{1/2}\chi)$ where
$K=-\Omega_KH_0^2$ is the curvature of the universe.  The null geodesic
equation in this metric is:
\begin{equation}
{d\over dr}\left( {d\nhat\over dr} \sin_K\chi \right) 
= - 2 {\partial\Psi\over\partial\nhat} \sin_K\chi.
\label{eq:gl}
\end{equation}
The initial conditions are $\nhat(\chi=0) = \nhat_0$ and
$\partial_\chi\nhat(\chi=0)=0$.

The usual method here is to apply the first-order Born approximation to
Eqs. (\ref{eq:gl}), i.e. we perform the integration over the unperturbed
photon trajectory. If we integrate forward, we find that:
\begin{equation}
\nhat(\chi) = {\bf n}_0 -2 \int_0^\chi W(\chi',\chi) 
{\partial\Psi(\chi',\nhat(\chi'))\over\partial\nhat} d\chi',
\end{equation}
where $W(\chi',\chi) = \cot_K\chi'-\cot_K\chi$. We may now apply the
second-order Born approximation, in which we integrate not over the
unperturbed photon trajectory but rather over the photon trajectory given
by the first-order Born approximation, Eq. (\ref{eq:gl}).  
Taylor-expanding the result to second order in $\Psi$ yields:
\begin{equation}
\nhat(\chi) = \nhat_0 -2 \int_0^\chi 
W(\chi',\chi)\partial_\nhat\Psi(\chi',\nhat_0) d\chi'
+ 4 \int_0^\chi \int_0^{\chi'} 
W(\chi'',\chi')W(\chi',\chi)
\partial^2_\nhat\Psi(\chi',\nhat_0)\cdot \partial_\nhat\Psi(\chi'',\nhat_0)
d\chi''d\chi'.
\end{equation}
The convergence and field rotation at radial coordinate $\chi$ are most
easily derived by taking the angular Fourier transform of this result.  
If we compute the deflection angle and perform the $(\kappa,\omega)$
decomposition of Eq. (\ref{eq:pdec}), we derive:
\begin{equation}
\kappa_\bfl = -l^2 \int_0^\chi W(\chi',\chi) \Psi_\bfl(\chi') d\chi'
-2 \sum_{\bfl'} (\bfl'\cdot\bfl)[\bfl'\cdot(\bfl-\bfl')] 
\int_0^\chi \int_0^{\chi'} W(\chi'',\chi')W(\chi',\chi)
\Psi_{\bfl'}(\chi') \Psi_{\bfl-\bfl'}(\chi'') d\chi''d\chi'
\label{eq:lensphi}
\end{equation}
and:
\begin{equation}
\omega_\bfl = -2\sum_{\bfl'} (\bfl'\cdot\star\bfl)[\bfl'\cdot(\bfl-\bfl')]
\int_0^\chi \int_0^{\chi'}
W(\chi'',\chi')W(\chi',\chi)\Psi_{\bfl'}(\chi') \Psi_{\bfl-\bfl'}(\chi'')
d\chi''d\chi'.
\label{eq:lensomega}
\end{equation}

We now turn our attention to the statistics of Eqs. (\ref{eq:lensphi}) and
(\ref{eq:lensomega}).  We assume that $\Psi$ can be described as a
Gaussian random field because even in the non-linear regime, our line of
sight passes through many regions of independent density fluctuation and
hence non-Gaussianity is suppressed by the Central Limit Theorem. The
power spectrum is:
\begin{equation}
\langle\Psi_{\bfl_1}(\chi_1)\Psi_{\bfl_2}(\chi_2)\rangle 
= \delta_{\bfl_1+\bfl_2,0}
{\cal C}_{l_1}^{\Psi\Psi}(\chi_1)\delta(\chi_1-\chi_2);
\end{equation}
here the projected potential power spectrum is determined using the Limber
equation:
\begin{equation}
{\cal C}_l^{\Psi\Psi}(\chi) = 
{1\over \sin_K^2\chi} P_\Psi\left( k={l\over\sin_K\chi},\chi\right) =
{9\sin_K^2\chi\over 4l^4} \Omega_{m0}^2H_0^4(1+z)^2 P_\delta\left( 
k={l\over\sin_K\chi},\chi\right).
\end{equation}
Here we have used the 3D power spectra of the Newtonian potential $\Psi$
and fractional density perturbation $\delta=\rho/\bar\rho-1$; these are
normalized in accordance with (here $q\in\{\Psi,\delta\}$):
\begin{equation}
P_q(k) = \int \langle q(0) q({\bf x}) \rangle 
e^{i{\bf k}\cdot{\bf x}} d^3{\bf x},
\end{equation}
so that the logarithmic band power is given by $\Delta_q^2(k) =
(k^3/2\pi^2) P(k)$. The lowest-order contribution to the convergence power
spectrum is given by the familiar result (here $\chi_0$ is the comoving
radial distance to the surface of last scatter):
\begin{equation}
C^{\kappa\kappa}_l = l^4 
\int_0^{\chi_0} W(\chi,\chi_0)^2 {\cal C}^{\Psi\Psi}_l(\chi) d\chi.
\label{eq:plus}
\end{equation}
(There are higher-order corrections to $C^{\kappa\kappa}_l$, but we do
not consider them here since the purpose of this paper is to investigate
lensing reconstruction, not to provide a precision theoretical
computation of the lensing power spectra.  Clearly if a sufficiently
high-precision measurement of $C_l^{\kappa\kappa}$ is made, higher-order
Born corrections should be considered in the theoretical interpretation
of the power spectrum.)
The field rotation power spectrum is given to lowest
order by: \begin{equation}
C^{\omega\omega}_l 
= 4 \sum_{\bfl'} (\bfl'\cdot\star\bfl)^2[\bfl'\cdot(\bfl-\bfl')]^2
\int_0^{\chi_0} d\chi\, \int_0^\chi d\chi'\; 
W(\chi,\chi_0)^2 W(\chi',\chi)^2 {\cal C}^{\Psi\Psi}_{l'}(\chi) {\cal
C}^{\Psi\Psi}_{\bfl-\bfl'}(\chi').
\label{eq:cross}
\end{equation}
Note that the lowest-order (in the Born expansion) contribution to
$C^{\omega\omega}_l$ comes from the trispectrum of the density field.  If
the density field is non-Gaussian and this non-Gaussianity is
insufficiently suppressed by the Central Limit Theorem, then
Eq. (\ref{eq:cross}) will also contain a term from the
connected trispectrum $\langle
\Psi\Psi\Psi\Psi\rangle_{connected}$.  However,
because the factor $W(\chi'',\chi')$ in Eq. (\ref{eq:lensomega}) vanishes
as
$\chi''\rightarrow\chi'$, it follows that the trispectrum components
contributing to $C^{\omega\omega}_l$ involve correlations between points
at widely spaced radial coordinates, which are suppressed.  (Conceptually,
this is because a single-screen lens only produces convergence and not
field rotation, regardless of its Gaussianity or lack thereof.  Thus if
structures at different radial distances are independent, as assumed 
in the Limber approximation, then there is no connected contribution to
$C^{\omega\omega}_l$.)

\subsection{Effect on lensing estimation}
\label{sec:effect}

\begin{figure}
\includegraphics[angle=-90,width=6.5in]{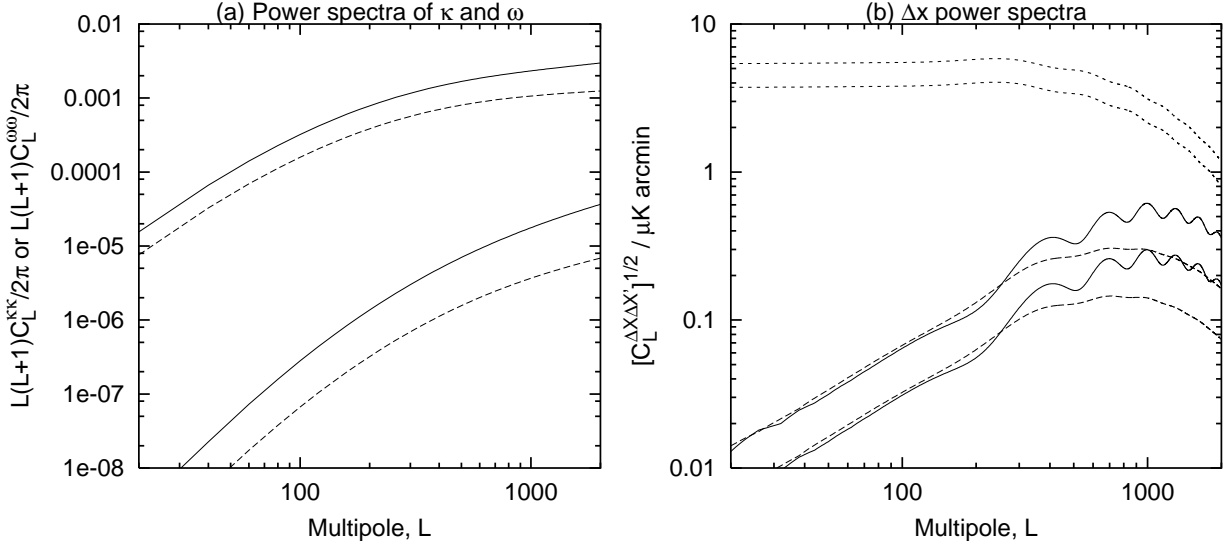}
\caption{\label{fig:omega}(a) The convergence (upper curves) and field
rotation (lower curves) power spectra in the fiducial cosmology.  These
are normalized to $\sigma_8^{\rm linear}=1.0$ (solid curves) and
$\sigma_8^{\rm linear}=0.7$ (dashed curves). (b) The power spectra of
$\Delta E$ (solid curves) and $\Delta B$ (long dashed) in ``noise units''
($\mu$K arcmin).  The short-dashed curves are the total $B$-mode power
introduced by the convergence component. The upper curves are calculated
for $\sigma_8^{\rm linear}=1.0$, the lower curves for $0.7$. }
\end{figure}

We have computed the field rotation power spectrum, Eq. (\ref{eq:cross})
for our fiducial cosmology using an analytic approximation to the growth
factor \cite{1992ARA&A..30..499C} and a nonlinear mapping of the power
spectrum \cite{1996MNRAS.280L..19P}. The results are plotted in Fig.
\ref{fig:omega}(a).

The effect of the field rotation on the lens reconstruction is to add an
additional term to the CMB given by $\Delta x = \nabla
x\cdot\star\nabla\Omega$.  The power spectrum of $\Delta x$ is given by:
\begin{eqnarray}
C^{\Delta T\Delta T}_l = && \frac{1}{\pi}
\sum_{\bfl'} {1\over {l'}^4} 
[\star\bfl'\cdot(\bfl-\bfl')]^2 C_{l'}^{\omega\omega} C_{\bfl-\bfl'}^{TT},
\nonumber \\
C^{\Delta T\Delta E}_l = && \frac{1}{\pi}
\sum_{\bfl'} {1\over {l'}^4} [\star\bfl'\cdot(\bfl-\bfl')]^2 
C_{l'}^{\omega\omega} C_{\bfl-\bfl'}^{TE} \cos 2\alpha,
\nonumber \\
C^{\Delta E\Delta E}_l = && \frac{1}{\pi}
\sum_{\bfl'} {1\over {l'}^4} [\star\bfl'\cdot(\bfl-\bfl')]^2 
C_{l'}^{\omega\omega} \left( C_{\bfl-\bfl'}^{EE} \cos^2 2\alpha
+ C_{\bfl-\bfl'}^{BB} \sin^2 2\alpha \right) ,
\nonumber \\
C^{\Delta B\Delta B}_l = && \frac{1}{\pi}
\sum_{\bfl'} {1\over {l'}^4} [\star\bfl'\cdot(\bfl-\bfl')]^2 
C_{l'}^{\omega\omega} \left( C_{\bfl-\bfl'}^{EE} \sin^2 2\alpha
+ C_{\bfl-\bfl'}^{BB} \cos^2 2\alpha \right),
\label{eq:cdx}
\end{eqnarray} 
where $\alpha = \phi_{\bfl'}-\phi_\bfl$. The field rotation is forbidden
to have first-order correlations with the primary CMB and the convergence
($C^{T\omega}_l = C^{E\omega}_l = C^{\kappa\omega}_l = 0$) by parity;
higher-order correlations with the primary CMB will be highly suppressed
because $\omega$ is determined by small-scale fluctuations in density
along the line of sight with window function that vanishes at the
last-scattering surface.  There are non-vanishing higher-order
correlations between $\kappa$ and $\omega$, but we do not investigate
these here.  [But note that by reducing the conditional covariance
$\langle \omega^2\rangle|_\Phi - (\langle\omega\rangle|_\Phi)^2$, these
correlations may enable us to reduce the ``noise'' due to field rotation
further.]

The $\Delta{\bf x}$ power spectrum [Fig. \ref{fig:omega}(b)] shows that
the $\omega$-induced modifications to the CMB $B$-modes are of the same
order as instrument noise when the latter is reduced to
$(N_l^{BB})^{1/2}\approx{\cal N}_P e^{l^2/2l_c^2}\approx 0.2$ $\mu$K
arcmin.  [Since we are trying to set $B_{\rm unlensed}=0$, contamination
in the $B$-modes is more serious for lensing than contamination in the
$E$-modes; this is made mathematically explicit by multiplication by
$\hat{\bf C}_{(0,0)}^{-1}$ in Eq. (\ref{eq:u}).] Since $\Delta{\bf x}$ has
vanishing first-order correlation with ${\bf x}$, one might conjecture
that the field rotation begins to interfere with lensing when the noise
${\cal N}_P$ is reduced to $\sim 0.2$ $\mu$K arcmin; however, $\Delta{\bf
x}$ is highly non-Gaussian and exhibits many higher-order correlations
with ${\bf x}$, so we should be cautious of trusting this conjecture.  In
the simulations (\S\ref{sec:numer1}), we find that even for our Reference
Experiment F with $0.25$ $\mu$K arcmin the field rotation does not
significantly contaminate the reconstruction of the convergence field --
it increases the mean squared error of the reconstruction by only $\sim$
15\%.  We conclude that (at least at the level of the experiments
considered here) the field rotation is not a problem for lens
reconstruction.

\section{Estimating the convergence power spectrum}
\label{sec:ape}

Having investigated the reconstruction of the lensing field, we turn our
attention to the convergence power spectrum, or equivalently the potential
power spectrum, since the two are related by $C^{\kappa\kappa}_l =
\frac{1}{4} l^4 C^{\Phi\Phi}_l$.  In this section, we will ignore any
complications associated with the field rotation as these are likely to be
small for near-term experiments.  In \S\ref{sec:cps}, we integrate the
likelihood function for the convergence to yield the ``grand likelihood
function'' for the lensing power spectrum; since this results in a
functional integral over lens realizations, we simplify the problem by
introducing a Gaussian approximation.  We make further approximations in
\S\ref{sec:estpow} to yield an estimator that is suitable for actual
computation.

\subsection{Likelihood function and Gaussian approximation}
\label{sec:cps}

Our basic approach, modeled after Ref. \cite{2003PhRvD..67d3001H}, is to
compute the grand likelihood function $\bar{\cal L}$, which is a function
of the lensing power spectrum:
\begin{equation}
\bar{\cal L}[C^{\kappa\kappa}_l] = 
-\ln \int {\cal D}\kappa\; e^{-\wp(\kappa)-{\cal L}(\kappa)}
=
\frac{1}{2} \ln\det {\bf C}^{\kappa\kappa}
-\ln \int {\cal D}\kappa\; \exp\left[
-\frac{1}{2} \kappa^\dagger{\bf C}^{\kappa\kappa\,-1}\kappa - {\cal L}(\kappa)
\right].
\label{eq:lint}
\end{equation}
The objective of this section is to develop formalism to compute the
minimum of $\bar{\cal L}$.  A ``practical'' version suitable for numerical
computation will be given in \S\ref{sec:estpow}.

The integral in Eq. (\ref{eq:lint}) has one dimension for each lensing
mode and hence cannot be performed by any brute-force technique.  In this
situation the preferred solution is usually to use a Markov chain;
unfortunately, the integral has of order $10^6$ dimensions, and the
integrand is expensive to compute, hence Eq. (\ref{eq:lint}) does not
appear to be solvable by Markov chains either. We therefore choose to
approximate Eq. (\ref{eq:lint}) as a Gaussian, in which case the
functional integral can be computed exactly:
\begin{equation}
\bar{\cal L}[C^{\kappa\kappa}_l]
\approx
\frac{1}{2} \ln\det {\bf C}^{\kappa\kappa}
+\frac{1}{2} \ln\det [{\cal F}(\kappa_\ast)+{\bf C}^{\kappa\kappa\,-1}]
+\frac{1}{2} \kappa_\ast^\dagger{\bf C}^{\kappa\kappa\,-1}\kappa_\ast 
+ {\cal L}(\kappa_\ast),
\end{equation}
where $\kappa_\ast$ is the point where ${\cal
L}+\frac{1}{2}\kappa^\dagger{\bf C}^{\kappa\kappa\,-1}\kappa$ is
minimized, and ${\cal F}(\kappa_\ast)$ is the curvature matrix, Eq.
(\ref{eq:curvaturematrix}), evaluated at the lens configuration
$g=(\kappa_\ast,0)$.

A grand likelihood gradient $\Gamma_l$ can then be defined:
\begin{eqnarray}
\Gamma_l = {\partial \bar{\cal L}\over\partial C^{\kappa\kappa}_l}
\approx &&
\frac{1}{2} \Tr\left[ {\bf C}^{\kappa\kappa\,-1} 
{\partial {\bf C}^{\kappa\kappa}\over \partial C^{\kappa\kappa}_l} \right]
+\frac{1}{2} \Tr\left[ ({\cal F}(\kappa_\ast)
+{\bf C}^{\kappa\kappa\,-1})^{-1} \left(
{ \partial{\cal F}(\kappa_\ast)\over\partial
C^{\kappa\kappa}_l} - 
{\bf C}^{\kappa\kappa\,-1}
{ \partial{\bf C}^{\kappa\kappa}\over\partial 
C^{\kappa\kappa}_l} {\bf C}^{\kappa\kappa\,-1} \right)
\right] \nonumber\\ && -
\frac{1}{2} \kappa_\ast^\dagger{\bf C}^{\kappa\kappa\,-1} 
{ \partial{\bf C}^{\kappa\kappa}\over\partial
C^{\kappa\kappa}_l} {\bf C}^{\kappa\kappa\,-1}\kappa_\ast
+ {\partial{\cal L}(\kappa_\ast)\over\partial C^{\kappa\kappa}_l}
\nonumber\\ && +
{\partial \kappa_\ast\over \partial C^{\kappa}_l} \left.\left\{
\Tr\left[ ({\cal F}(\kappa)+{\bf C}^{\kappa\kappa\,-1})^{-1} 
{ \partial{\cal F}(\kappa)\over\partial \kappa} \right]
+ {\partial\over\partial\kappa} \left(
\frac{1}{2} \kappa^\dagger{\bf C}^{\kappa\kappa\,-1}\kappa + {\cal L}(\kappa)
\right) \right\} \right|_{\kappa=\kappa_\ast}.
\label{eq:gamma0}
\end{eqnarray}
Note that ${\cal L}(\kappa_\ast)$, and hence ${\cal F}(\kappa_\ast)$, do
not depend on $C^{\kappa\kappa}_l$ except implicitly through
$\kappa_\ast$. Also, if we use that $\kappa_\ast$ is the maximum of
$\frac{1}{2} \kappa^\dagger{\bf C}^{\kappa\kappa\,-1}\kappa + {\cal L}$
with respect to $\kappa$, we find that the final derivative with respect
to $\kappa$ in this equation vanishes.  We further note that $\partial{\bf
C}^{\kappa\kappa}/\partial C^{\kappa\kappa}_l$ is simply the projection
operator onto the $l$ representation of $SO(3)$, i.e. in harmonic space it
has 1's as diagonal elements with multipole $l$ and 0's everywhere else.
Defining $d_l$ to be the number of modes of multipole $l$ (note that on
the sphere, $d_l=2l+1$), this allows us to simplify Eq. (\ref{eq:gamma0})
to:
\begin{equation}
\Gamma_l \approx \frac{d_l}{2 C_l^{\kappa\kappa}} -
\sum_{\bfl:\,|\bfl|=l} \frac{
[({\cal F}(\kappa_\ast)+{\bf C}^{\kappa\kappa\,-1})^{-1}]_{\bfl,\bfl} }
{2C_l^{\kappa\kappa\,2}}
- \sum_{\bfl:\,|\bfl|=l}
{|\kappa_{\ast\,\bfl}|^2\over 2 C^{\kappa\kappa\,2}_l}
+ {\partial \kappa_\ast\over \partial C^{\kappa\kappa}_l}
\Tr\left[ ({\cal F}(\kappa_\ast)+{\bf C}^{\kappa\kappa\,-1})^{-1} \left.
{ \partial{\cal F}(\kappa)\over\partial 
\kappa} \right|_{\kappa_\ast} \right].
\label{eq:gamma}
\end{equation}
Here the sums are over all modes $\bfl$ of multipole $l$.

It sometimes occurs that we wish to estimate the lensing power spectrum
not by estimating the individual $C_l^{\kappa\kappa}$, but rather by
``binning'' the power spectrum.  This is useful if, e.g. the $(S/N)^2$ per
multipole is low or if the partial-sky nature of a survey causes confusion
between power in neighboring multipoles.  In this case, we introduce
``basis functions'' $\{{\cal M}^\mu\}$ for the lensing power spectrum:
\begin{equation}
C_l^{\kappa\kappa} = \sum_\mu c_\mu {\cal M}_l^\mu;
\label{eq:mmu}
\end{equation}
the coefficients $c_\mu$ are now to be estimated.  The maximum-likelihood
estimator is then the choice of $c_\mu$ that satisfies:
\begin{equation}
\sum_l {\cal M}_l^\mu \Gamma_l = 0 \;\;\;\forall \mu.
\label{eq:mmuest}
\end{equation}

\subsection{Practical estimator and uncertainty}
\label{sec:estpow}

Ideally, we would like to implement the maximum-likelihood estimator for
the coefficients $c_\mu$, i.e. Eq. (\ref{eq:mmuest}).  Unfortunately, this
involves setting to zero some linear combination of the $\Gamma_l$'s given
by Eq. (\ref{eq:gamma}), which is a highly non-trivial task. We therefore
take the approximation that the curvature matrix ${\cal F}$ does not
depend on $\kappa$, then $\Gamma_l$ is seen to depend only on the
quantities $C^{\kappa\kappa}_l$ and:
\begin{equation}
\dot v_l \equiv {1\over 2 C^{\kappa\kappa\,2}_l} 
\sum_{\bfl:\,|\bfl|=l} |\kappa_{\ast\,\bfl}|^2,
\label{eq:vl}
\end{equation}
which explicitly depends on $C^{\kappa\kappa}_l$ but is also implicitly a
function of $C^{\kappa\kappa}_l$ through its dependence on $\kappa_\ast$.
Note that the functional form of $\Gamma_l$ is
$\Gamma_l(C^{\kappa\kappa}_l,\dot v_l) =
\Gamma_l^{(0)}(C^{\kappa\kappa}_l)-\dot v_l$.  Eq. (\ref{eq:mmuest}) then
reads:
\begin{equation}
0= \sum_l {\cal M}_l^\mu
\left[  \Gamma_l^{(0)} (C^{\kappa\kappa}_l) - \dot v_l(c_\nu) \right]
= \Gamma_\mu^{(0)}(c_\nu) - v_\mu(c_\nu),
\end{equation}
where we have defined $v_\mu(c_\nu) = \sum_l {\cal M}_l^\mu \dot
v_l(c_\nu)$. We are thus attempting to solve $v_\mu(c_\nu) =
\Gamma_\mu^{(0)}$, but $\Gamma_\mu^{(0)}$ is some complicated function of
the convergence power spectrum coefficients $\{c_\nu\}$.  We solve this
problem by approximating $\Gamma_\mu^{(0)}(c_\nu) \approx \langle
v_\mu(c_\nu)\rangle_{LSS[c_\nu]}$, i.e. the expected value of
$v_\mu(c_\nu)$ where the LSS realizations are drawn from a lensing
convergence power spectrum $C^{\kappa\kappa}_l = \sum_\nu c_\nu{\cal
M}_l^\nu$.  We therefore use the estimator:
\begin{equation}
v_\mu(c_\nu) = \langle v_\mu(c_\nu)\rangle_{LSS[c_\nu]}\;\;\;\forall\mu.
\label{eq:v}
\end{equation}

Eq. (\ref{eq:v}) is somewhat abstract, so we clarify its meaning here.  
The statistic $v_\mu(c_\nu)$ is proportional to the power spectrum of the
iterative convergence estimator obtained by solving Eq. (\ref{eq:estcon});
this depends on the prior power spectrum $C^{\kappa\kappa}_l = \sum_\nu
c_\nu{\cal M}_l^\nu$ as well as on the data.  The solution $\{c_\nu\}$ to
Eq. (\ref{eq:v}) is the set of power spectrum coefficients for which
$v_\mu$ equals its expected value (which is most easily determined via
Monte Carlo simulation).  This approach has the advantage of ``calibrating
out'' the noise biases discussed by Ref. \cite{2003astro.ph..2536K}. (Note
that some convergence modes -- those corresponding to large eigenvalues of
the curvature ${\cal F}$ -- are reconstructed better than others.  What is
especially useful about $v_\mu$, or equivalently the power spectrum of the
iterative estimator, is that the iterative estimator filters out the
poorly reconstructed modes.  Thus the convergence modes that are
reconstructed more accurately are weighted more heavily in determining
$v_\mu$ and hence in determining the convergence power spectrum.)

Finally, we wish to determine the uncertainty on the solution $\{c_\nu\}$
to Eq. (\ref{eq:v}).  If we average over many convergence modes, then this
uncertainty should be given by the inverse of the grand Fisher matrix
$^{(G)}{\bf F}$ for power spectrum determination:
\begin{equation}
^{(G)}F_{\mu,\mu'} = \langle \Gamma_\mu \Gamma_{\mu'} \rangle_{LSS}
\approx \langle \delta v_\mu \delta v_{\mu'} \rangle_{LSS}
= \langle v_\mu v_{\mu'} \rangle_{LSS} - \langle v_\mu \rangle_{LSS}
\langle v_{\mu'} \rangle_{LSS},
\label{eq:grandfisher}
\end{equation}
i.e. $^{(G)}{\bf F}$ is the covariance matrix of $v_\mu$.  If the {\em
reconstructed} convergence $\kappa_\ast$ can be approximated as a Gaussian
random field (which is true in the case where the reconstruction has high
signal-to-noise ratio since in this case the $\kappa_\ast\approx\kappa$,
which is Gaussian because $\kappa$ is produced by many LSS fluctuations
along the line of sight), then we can take the Gaussian approximation to
Eq. (\ref{eq:grandfisher}).  This is obtained by considering the
covariance of $\dot v_l$ according to Eq. (\ref{eq:vl}) and using Wick's
theorem; this yields:
\begin{equation}
^{(G)}F_{\mu,\mu'} = \sum_l {d_l {\cal M}^\mu_l {\cal M}^{\mu'}_l \over 2}
\left( \frac{C^{\kappa_\ast\kappa_\ast}_l}{ C^{\kappa\kappa\,2}_l}\right)^2.
\label{eq:gfgauss}
\end{equation}
We remind the reader once again that the approximation Eq.
(\ref{eq:gfgauss}) to the power spectrum estimation uncertainty is only
valid if the reconstructed convergence field $\kappa_\ast$ is
approximately Gaussian.  If $\kappa_\ast$ has a significant trispectrum
when averaged over LSS+CMB+noise realizations, then Eq.
(\ref{eq:grandfisher}) must be used instead.  This is only a problem in
the low signal-to-noise (high $l$) regime in which lensing modes cannot be
reconstructed individually and their power is only statistically detected.

\section{Numerical simulations}
\label{sec:numer}

\begin{figure}
\includegraphics[angle=-90,width=7in]{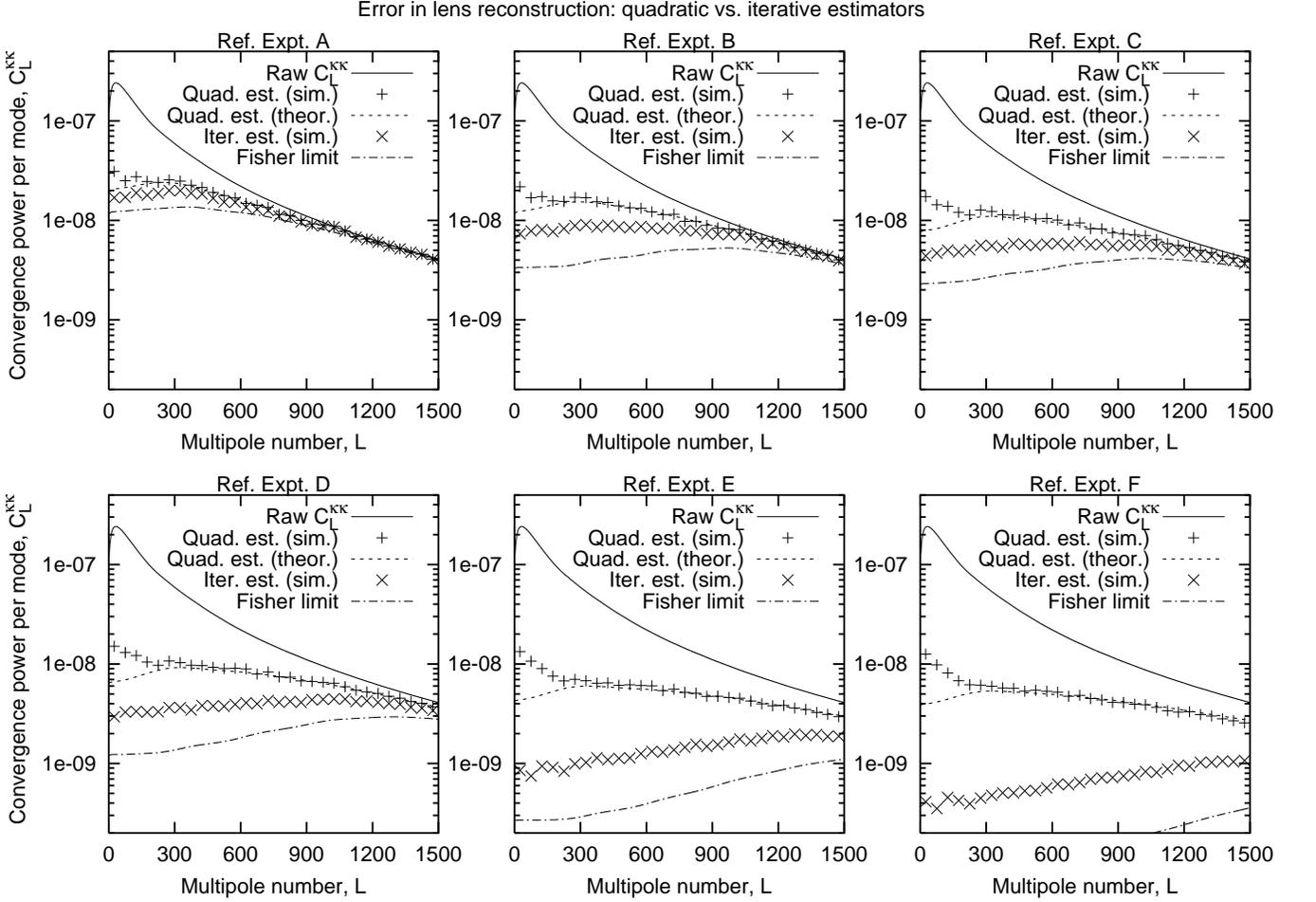}
\caption{\label{fig:ckappa}The power spectrum of the error in the
convergence reconstruction for Reference Expts. A--F.  The top curve in
each panel shows the overall convergence power spectrum
$C^{\kappa\kappa}_l$.  The middle curve shows the theoretical, i.e. from
Eq. (\ref{eq:qfisher}) power spectrum of the convergence error
$\hat\kappa-\kappa$ in the Wiener-filtered quadratic estimator Eq.
(\ref{eq:wfquad}); the ``$+$'' data points indicate the power spectrum of
this error as recovered from simulations.  The error power spectrum for
the iterative estimator Eq. (\ref{eq:iter}), again as recovered from
simulations, is shown with the ``$\times$'' data points. The bottom curve
shows the theoretical best performance if the Fisher matrix limit Eq.
(\ref{eq:fphi}) can be achieved, i.e. if we had a truly optimal estimator
and no curvature corrections. Note the more dramatic improvement provided
by the iterative estimator when the noise is small.  Field rotation was
neglecting in the calculations for this figure. } 
\end{figure}

\begin{figure}
\includegraphics[angle=0,width=6.9in]{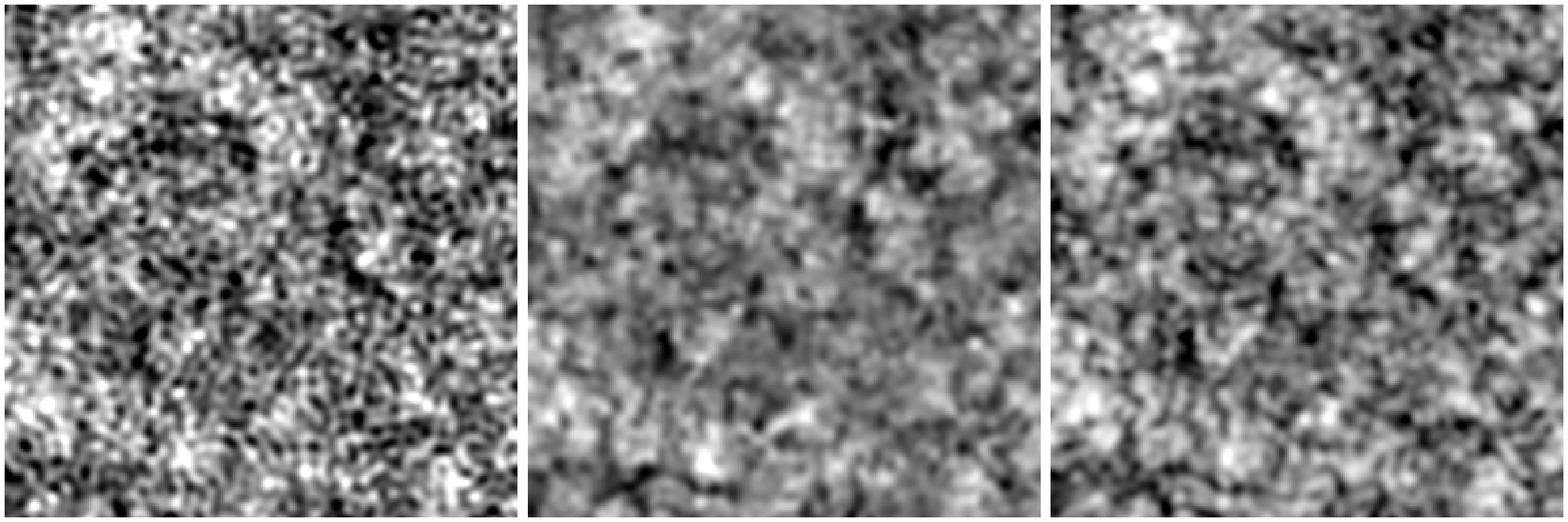}
\caption{\label{fig:refc}A simulated reconstruction of the lensing
convergence using polarization and Reference Expt. C parameters.  In the
left panel, we display the realization of the convergence field $\kappa$
used to produce the simulated CMB.  The reconstructions using the
Wiener-filtered quadratic estimator and the iterative estimator are shown
in the center and right panels, respectively.  These frames are each
$8^\circ 32'$ in angular width, corresponding to 1/16 of the simulated
area; the scale ranges from black (diverging, $\kappa=-0.12$) through
white (converging, $\kappa=+0.12$).  Although all lensing multipoles up to
$l=3600$ are simulated, we have only displayed the $l\le 1600$ modes in
these figures for clarity. Field rotation was neglected in the
calculations for this figure.} 
\end{figure}

Throughout our derivation of lensing and tensor power estimators, we have
made various approximations that should be tested.  The most robust way to
do this is to conduct a numerical simulation of the CMB and lensing field,
and then construct lensing estimators, comparing the error to the
theoretical estimates of Eqs. (\ref{eq:fphi}) and (\ref{eq:qfisher}). In
all cases, we have used a flat sky with toroidal boundary conditions.  We
will only simulate the CMB polarization here; formally, the
polarization-only estimators are obtained by setting ${\cal N}_T=\infty$
in the relevant equations.

\subsection{Reconstructing the convergence}
\label{sec:numer1}

The simplest simulations involve reconstruction of the convergence
$\kappa$.  We generate simulated CMB $T$, $Q$, $U$, and $\kappa$ fields on
a $34^\circ 08'$ square patch of sky with resolution $1$ arcmin per pixel
($2048\times 2048$ pixels); lens the simulated CMB; and add appropriate
noise.

We wish to compare the quadratic estimator, Eq. (\ref{eq:wfquad}) with our
new estimator, Eq. (\ref{eq:estcon}).  The former is relatively
straightforward to compute; the latter requires that we apply the methods
of \S\ref{sec:estcon}.  We simulate Gaussian random realizations of the
$Q$, $U$, and $\kappa$ fields, perform the lensing re-mapping, and add
appropriate noise.  We then compute the estimators of Eqs.
(\ref{eq:wfquad}) and (\ref{eq:estcon}).  There are two tricks that are
very useful in numerical computation of these estimators: first,
simultaneous computation of inner products ${\bf
t}^\dagger\sigma^\kappa_\bfl{\bf u}$ for all $\bfl$; and second,
stochastic trace computation.  We discuss each of these here.

The simultaneous computation of inner products was introduced by Refs.
\cite{2001ApJ...557L..79H, 2002ApJ...574..566H} in order to compute
quadratic estimators.  A general version of this is (on a flat sky; see
Ref. \cite{2003PhRvD..67h3002O} for an all-sky version):
\begin{equation}
\sum_\bfl \left({l^2\over 2}\right)
({\bf t}^\dagger\sigma^\kappa_{-\bfl}{\bf u}) e^{i\bfl\cdot\nhat}
= \sum_{X\in\{T,Q,U\}} \nabla_\nhat\cdot
[t_X^\ast(\nhat)\nabla_\nhat u_X(\nhat)].
\label{eq:tutrick}
\end{equation}
(Note that this equation requires that ${\bf t}$ and ${\bf u}$ be written
in the $\{T,Q,U\}$ basis since $E$ and $B$ have different transformation
properties under lensing. Also the asterisk on $t_X$ is of course
unnecessary if ${\bf t}$ is a real field.) If ${\bf t}$ and ${\bf u}$ are
expressed in real-space, then the right-hand side is easily evaluated.  
The quantities ${1\over 2}l^2{\bf t}^\dagger\sigma^\kappa_\bfl{\bf u}$ are
then obtained via a fast Fourier transform; division then trivially
removes the $l^2/2$.  There is a zero-wavenumber mode corresponding to
$l=0$ which presents a problem for division.  Here we simply set this
convergence mode to zero; in the complete all-sky treatment this would be
justified by noting that the convergence is $-{1\over 2}$ times the
divergence of the deflection angle vector, hence $\int\kappa\,d^2\nhat=0$
and so $\kappa_{l=0}=0$.  [There is a corresponding trick for the field
rotation: Eq. (\ref{eq:tutrick}) remains valid if we make the replacements
$\sigma^\kappa_\bfl\rightarrow\sigma^\omega_\bfl$ and $\nabla_\nhat
u_X(\nhat)\rightarrow\star\nabla_\nhat u_X(\nhat)$.]

The trace in Eq. (\ref{eq:estcon}) is most easily evaluated
stochastically: if we generate a random noise vector $\eta$ with
covariance ${\bf N}$, then the trace is equal to the expectation value:
\begin{equation}
\Tr\left[\Lambda_g^{\dagger\,-1}{\bf C}_{(0,0)}^{-1}
\sigma^\kappa_\bfl {\bf C}
\hat{\bf C}_{(0,0)}^{-1} \Lambda_g^{-1}{\bf N}\right]
= \langle (\hat{\bf C}_{(0,0)}^{-1} \Lambda_g^{-1} \eta)^\dagger 
\sigma^\kappa_\bfl {\bf C}
\hat{\bf C}_{(0,0)}^{-1} \Lambda_g^{-1} \eta \rangle.
\label{eq:stochastictrace}
\end{equation}
If this Monte Carlo method is used to compute the trace, then the Monte
Carlo error in its computation for one realization of $\eta$ is less than
or equal to the instrument noise contribution to the uncertainty on the
right-hand side of Eq. (\ref{eq:estcon}).  [This is because the right-hand
side of Eq. (\ref{eq:estcon}) is a quadratic function of $\hat{\bf x}$,
with covariance $\hat{\bf C}$, which is greater than the noise covariance
${\bf N}$ along all directions.] Since the Monte Carlo error variance
scales as the reciprocal of the number of realizations of $\eta$ used, it
follows that of order a few realizations of $\eta$ are sufficient in
evaluating Eq. (\ref{eq:stochastictrace}).  In fact for the Reference
Experiments described here, we find that there is little gain in taking
more than one realization of $\eta$.

\begin{figure}
\includegraphics[angle=-90,width=6.9in]{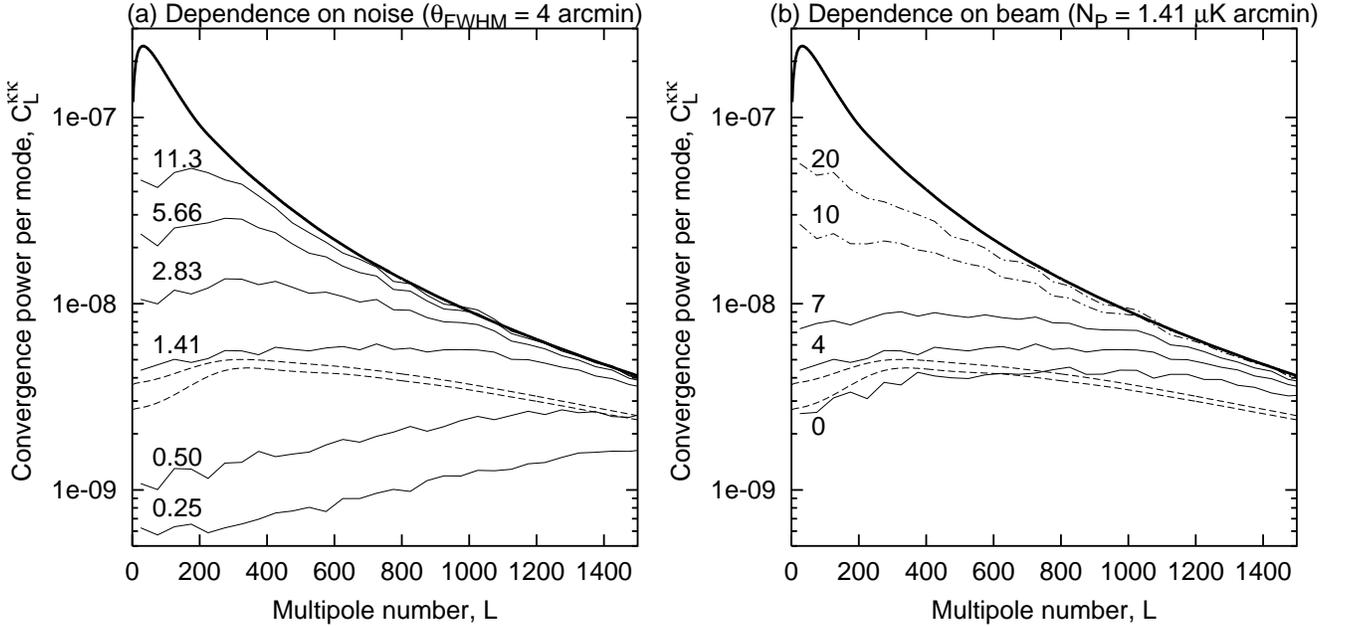}
\caption{\label{fig:ckappa2}The dependence of the mean squared error in
lensing reconstruction, $\langle |\kappa_\bfl-\hat\kappa_\bfl|^2\rangle$,
on the instrument parameters.  The baseline is Ref. Expt. C, ${\cal
N}_P=1.41$ $\mu$K arcmin, $\theta_{FWHM}=4$ arcmin.  The thick solid line
is the raw power spectrum $C^{\kappa\kappa}_l$; the thin solid lines
indicate the mean squared error for the lensing reconstruction using the
iterative estimator.  As described in the text, the iterative estimator is
unusable for wide-beam experiments ($\ge$10 arcmin); we used the quadratic
estimator for these cases (dot-dashed curves). The dashed lines indicate
the ideal zero-noise reconstruction error from the quadratic estimator
according to Eq. (\ref{eq:qfisher}) with polarization only (top) and
temperature+polarization (bottom). (a) Changing ${\cal N}_P$; units are
$\mu$K arcmin.  (b) Changing $\theta_{FWHM}$; units are arcmin. }
\end{figure}

We solve Eq. (\ref{eq:estcon}) using the iterative procedure:
\begin{equation}
\kappa_{n+1,\bfl}^\ast = \zeta_l
[\Lambda_g^{-1}]_{-\bfl,-\bfl'} C^{\kappa\kappa}_{l'} \left\{
(\hat{\bf C}_{(0,0)}^{-1} \Lambda_g^{-1}\hat{\bf x})^\dagger
\sigma^\kappa_{\bfl'} {\bf C}
\hat{\bf C}_{(0,0)}^{-1} \Lambda_g^{-1}\hat{\bf x}
- \Tr\left[ \Lambda_g^{\dagger\,-1}\hat{\bf C}_{(0,0)}^{-1}
\sigma^\kappa_{\bfl'} {\bf C}
\hat{\bf C}_{(0,0)}^{-1} \Lambda_g^{-1}
{\bf N} \right] \right\}
+(1-\zeta_l) \kappa_{n,\bfl}^\ast.
\label{eq:iter}
\end{equation}
Here $g$ is the lens configuration with convergence $\kappa$ and no
rotation: $g=(\kappa,0)$, and the $\zeta_l$ are convergence parameters; we
choose them to be:
\begin{equation}
\zeta_l = \frac{\zeta_{(c)}}{1+C^{\kappa\kappa}_lF^{\kappa\kappa}_l},
\end{equation}
where $\zeta_{(c)}$ is a constant satisfying $0<\zeta_{(c)}<2$. It is
found that the Wiener-filtered quadratic estimator, Eq. (\ref{eq:wfquad}),
is a good choice for initializing this iteration. The choice of
$\zeta_{(c)}$ is an intricate issue: if it is set too small, the rate of
convergence of the iteration becomes unacceptably slow; if it is set too
high, the iteration can fail to converge entirely. The convergence can be
understood by approximating Eq. (\ref{eq:iter}) as linear in $\kappa_n$:
\begin{equation}
\kappa_{n+1,\bfl}^\ast \approx \zeta_l
C_l^{\kappa\kappa}[{\cal F}(\kappa_n-\hat\kappa)]_\bfl +
(1-\zeta_l)\kappa_{n,\bfl};
\end{equation}
here we have approximated the response matrix of the likelihood gradient
using the curvature matrix.  Then the requirement for convergence is that
all of the (possibly complex) eigenvalues of the matrix:
\begin{equation}
R_{\bfl,\bfl'} = \delta_{\bfl,\bfl'} - \zeta_{(c)} {
C_{\bfl,\bfl'}^{\kappa\kappa\,-1}+ {\cal F}_{\bfl,\bfl'} \over
C_l^{\kappa\kappa\,-1} +  F_l^{\kappa\kappa} }
\label{eq:rll}
\end{equation}
lie within the unit circle.  Note that, averaged over CMB+noise
realizations, $\langle{\cal F}\rangle = {\bf F}$, and hence $\langle{\bf
R}\rangle=(1-\zeta_{(c)}){\bf 1}$; hence we conclude that the iterative
procedure should be convergent for $0<\zeta_{(c)}<2$ in the absence of
curvature corrections.  In reality, very small values of $\zeta_{(c)}$ may
be necessary for convergence, especially in cases where curvature
corrections are large.  Since ${\cal F} + {\bf C}^{\kappa\kappa\,-1}$ is
positive definite at the maximum posterior probability point, there is
always a positive value of $\zeta_{(c)}$ that results in convergence. The
cases in which the small values of $\zeta_{(c)}$ are required are those in
which curvature corrections are large; we have found from our simulations
that these are the low-noise experiments.  Convergence is generally found
to be faster for the high $l$ convergence modes.

One problem we have encountered is that for experiments with low noise and
wide beam ($\theta_{FWHM}\ge 10$ arcmin), the iterative estimator given by
Eq. (\ref{eq:iter}) is unstable.  This instability arises because the
noise $N_l$ is strongly blue; hence the de-lensing operation
$\Lambda_g^{-1}$ in Eq. (\ref{eq:iter}) mixes high-multipole noise down to
lower multipoles where it disrupts the lensing estimation.  This problem
is in principle solvable by using the correct $({\bf C}+\Lambda_g^{-1}{\bf
N}\Lambda_g^{\dagger\,-1})^{-1}$ weight function in place of ${\bf
C}_{(0,0)}^{-1}$ in Eq. (\ref{eq:ape2b}).  However, since this occurs in a
regime where the iterative approach does not improve upon the quadratic
estimator approach anyway, we recommend simply using the quadratic
estimator for wide-beam experiments.

\begin{figure}
\includegraphics[angle=-90,width=6in]{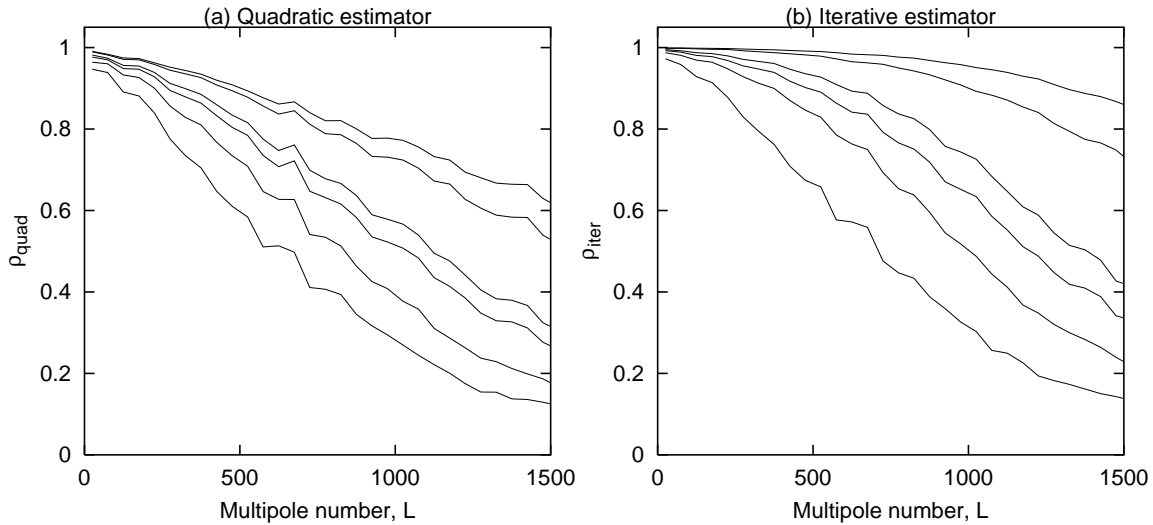}
\caption{\label{fig:fcorr}The correlation coefficient $\rho = \langle\kappa\hat\kappa\rangle /
\langle\kappa^2\rangle^{1/2} \langle\hat\kappa^2\rangle^{1/2}$ between the estimated and reconstructed lensing
convergences as a function of multipole $l$, as determined in simulations.  The correlation coefficients for the quadratic
estimator are shown in panel (a); those for the iterative estimator are shown in panel (b).  In both of these panels, the eight
curves are for Reference Expts. A, B, C, D, E, and F (bottom to top) from Table \ref{tab:expt}.  Field rotation was not
included in the calculations for this figure.
}
\end{figure}

We illustrate by considering the reconstruction of lensing using Reference
Experiments A--F.  The residual error in the reconstructed convergence
$\hat\kappa$, as measured by computing the power spectrum of the
difference $\hat\kappa-\kappa$ between input and reconstructed convergence
maps, is shown in Fig. \ref{fig:ckappa} for both quadratic and iterative
estimators.  For the iterative estimator applied to Ref. Expt. C, we set
$\zeta_{(c)}=0.12$ in Eq. (\ref{eq:iter}), used three realizations of the
$\eta$ field in Eq. (\ref{eq:stochastictrace}), and performed 64
iterations.  Ref. Expt. F has a lower noise level and so it is necessary
to use the smaller convergence parameter $\zeta_{(c)}=0.04$; the
convergence is thus slower and we used 256 iterations. Ref. Expt. A has a
higher noise level and so we can use $\zeta_{(c)}=0.2$ and 24 iterations.  
Maps of the input and reconstructed convergence fields for the Ref. Expt.
C reconstruction are shown in Fig. \ref{fig:refc}.  The dependence of the
iterative estimator reconstruction accuracy on noise ${\cal N}_P$ and beam
size (FWHM) $\theta_{FWHM}$ is shown in Fig. \ref{fig:ckappa2}. We have
also displayed in Fig. \ref{fig:ckappa2} the (theoretical) reconstruction
error curves for the quadratic estimator in the absence of instrument
noise.  These curves represent the fundamental limit to the reconstruction
accuracy possible with quadratic estimators; it is readily seen that the
iterative estimator can do better if noise is low (${\cal N}_P<0.5$--$1$
$\mu$K arcmin, depending on the range of $l$ considered). [Note that we
display $C^{\kappa\kappa}_l$ in these plots, whereas some authors have
displayed instead $l(l+1)C^{\bf dd}/2\pi$, where ${\bf d}=\nabla\Phi$ is
the deflection angle.  The two are related by $l(l+1)C^{\bf dd}/2\pi =
(2/\pi)C^{\kappa\kappa}_l$.]

The accuracy of reconstruction can also be represented by the correlation
coefficient $\rho_l = C^{\kappa\hat\kappa}_l/\sqrt{ C^{\kappa\kappa}_l
C^{\hat\kappa\hat\kappa}_l}$. The correlation coefficient is the figure of
merit if the objective is to cross-correlate the convergence from CMB
lensing with another tracer of the density (e.g. weak lensing of
galaxies), since the signal-to-noise ratio of the cross-correlation is
determined by $\rho_l$. We have plotted the correlation coefficient in
Fig. \ref{fig:fcorr} for the various Reference Experiments.  The iterative
estimator offers improved reconstruction, especially for the lower-noise
experiments (C--F).

\begin{figure}
\includegraphics[angle=-90, width=6.5in]{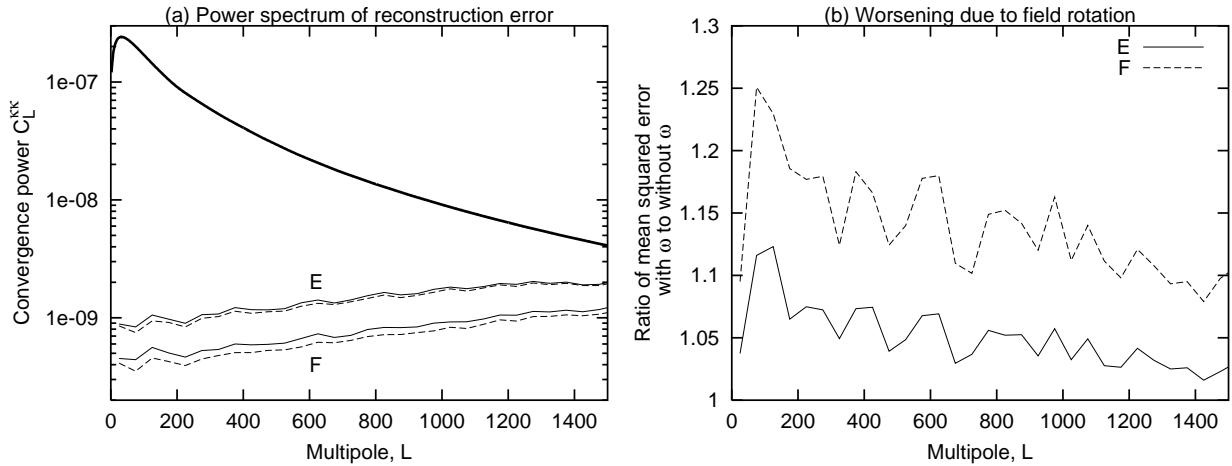}
\caption{\label{fig:omegaeff}The effect of field rotation on lensing
estimation for Ref. Expts. E and F.  (a) Power spectra of the convergence
$\kappa$ (thick solid line), convergence error $\kappa-\hat\kappa$ with
field rotation (thin solid lines), and convergence error
$\kappa-\hat\kappa$ without field rotation (thin dashed lines).  (b) The
worsening of the reconstruction due to the presence of field rotation, as
measured by the ratio of power spectra of the convergence errors:
$\langle|(\kappa-\hat\kappa)_l|^2\rangle_{{\rm
with}\;\omega}/\langle|(\kappa-\hat\kappa)_l|^2\rangle_{{\rm
without}\;\omega}$.  The same CMB, LSS, and noise (scaled appropriately to
the experiment) realizations were used for all the simulations in this
figure.}
\end{figure}

Up until this point we have neglected the field rotation $\omega$; we
should verify that this is justified.  We do this by introducing field
rotation with power spectrum given by Eq. (\ref{eq:cross}) as computed in
\S\ref{sec:effect} with normalization $\sigma_8^{\rm linear}=0.84$.  We
then compare the performance of the iterative estimator, Eq.
(\ref{eq:iter}), with and without the field rotation.  The comparison is
shown in Fig. \ref{fig:omegaeff}; it is seen that the field rotation
increases the mean squared error of the reconstruction by only $\sim$ 5\%
for Ref. Expt. E (0.5 $\mu$K arcmin noise, 2 arcmin beam) and $\sim$ 15\%
for Ref. Expt. F (0.25 $\mu$K arcmin noise, 2 arcmin beam).

As a final note, we find that for low noise levels, a large number of
iterations is required because our iterative process is ill-conditioned.  
Indeed, it is possible that there are eigenvalues of ${\bf R}$ that are so
close to unity that their corresponding modes have not converged even
after tens or hundreds of iterations; if this is the case, then it should
be possible to improve upon our results by increasing the number of
iterations, or by finding an iterative scheme that converges faster than
Eq. (\ref{eq:iter}). This is allowed by the Fisher matrix noise limits,
which are significantly lower than the achieved noise levels (see Fig.
\ref{fig:ckappa}). We consider this possibility unlikely since we tried
increasing the number of iterations in several of the simulations and
found little improvement.  Additionally, modes with eigenvalue
$\lambda_{\bf R}$ close to unity correspond to flat directions of the
curvature matrix ${\cal F}$ [see Eq. (\ref{eq:rll})]; such directions,
however, cannot be reconstructed accurately regardless of how many
iterations are used.

\subsection{Extracting the convergence power spectrum}

We compute the lensing power spectrum from simulated data by solving Eq.
(\ref{eq:v}).  The approach, once again, is iterative: we adjust the power
spectrum $C^{\kappa\kappa}_l$ until $v_\mu=\langle v_\mu\rangle_{LSS}$.  
[Note that both the left and right sides of Eq. (\ref{eq:v}) depend on
$C^{\kappa\kappa}_l$.] We will attempt here to compute the binned power
spectrum, i.e. we choose a basis for the convergence power spectrum given
by:
\begin{equation}
{\cal M}_\mu^l = \left\{ \begin{array}{lcl}
1 & & \mu\Delta l< l \le (\mu+1)\Delta l, \\
0 & & {\rm otherwise},
\end{array} \right.
\label{eq:mubasis}
\end{equation}
where $\Delta l$ is the bin width and $\mu$ ranges from 0 through
$N_{bin}-1$.  We are thus attempting to reconstruct the power spectrum in
$N_{bin}$ bins, equally spaced out to maximum multipole $l_{\rm
max}=N_{bin}\Delta l$.

Our iterative algorithm for solving Eq. (\ref{eq:v}) is:

\begin{equation}
c_{\mu,n+1} = c_{\mu,n}^{1-\zeta_{(p)}} \left[ c_{\mu,n} +
\frac{d_\mu}{2 \langle v_\mu(c_{\nu,n})\rangle_{LSS[c_{\nu,n}]}}
\left( \frac{ v_\mu(c_{\nu,n})} 
{\langle v_\mu(c_{\nu,n})\rangle_{LSS[c_{\nu,n}]} }
- 1\right) \right]^{\zeta_{(p)}},
\label{eq:poweriter}
\end{equation}
where $n$ represents the iteration number, and $d_\mu$ is the number of
modes that fall into the $\mu$th band. It is readily apparent that the
final values $c_{\nu,\infty}$ will satisfy $v_\mu = \langle
v_\mu\rangle_{LSS}$.

In order to compute the convergence power spectrum estimator, the expected
value $\langle v_\mu(c_{\nu,n})\rangle_{LSS[c_{\nu,n}]}$ must be
determined; the simplest method for doing this is via Monte Carlo
simulations. Since in the end we are solving the equation $v_\mu = \langle
v_\mu\rangle_{LSS}$, we want to make sure that the Monte Carlo-induced
error in the right-hand side of this equation is small compared with the
statistical error in the left-hand side (which depends only on the data
and on $c_{\nu,n}$).  It is trivial to see that after $N_{MC}$ Monte Carlo
simulations, the variance in determination of the right-hand side is
$1/N_{MC}$ of the statistical variance in the left-hand side. Therefore,
we expect that if we use $N_{MC}$ Monte Carlo simulations to determine
$\langle v_\mu\rangle_{LSS}$, then the variance of our determination of
the convergence power spectrum will increase by a factor of $1+1/N_{MC}$.
A reasonable choice, then, is to take $N_{MC}=3$, which results in 15\%
increase in the variance of the power spectrum estimator over the case of
$N_{MC}=\infty$ (exact computation of $\langle v_\mu\rangle_{LSS}$).  The
uncertainty in the power spectrum estimation can then be estimated from
Eq. (\ref{eq:grandfisher}) with the correction for $N_{MC}$:
\begin{equation}
\sigma_{c_\mu} = {C^{\kappa\kappa\,2}_l\over C^{\kappa_\ast\kappa_\ast}_l}
\sqrt{{2\over d_\mu}(1+N_{MC}^{-1})}.
\label{eq:gf-sigma}
\end{equation}
(Note that this is the standard Gaussian formula for error bars, except
that it is corrected for $N_{MC}$ and is written in terms of the filtered
power spectrum $C^{\kappa_\ast\kappa_\ast}_l$ instead of the noise power.)

\begin{figure}
\includegraphics[angle=-90,width=6.5in]{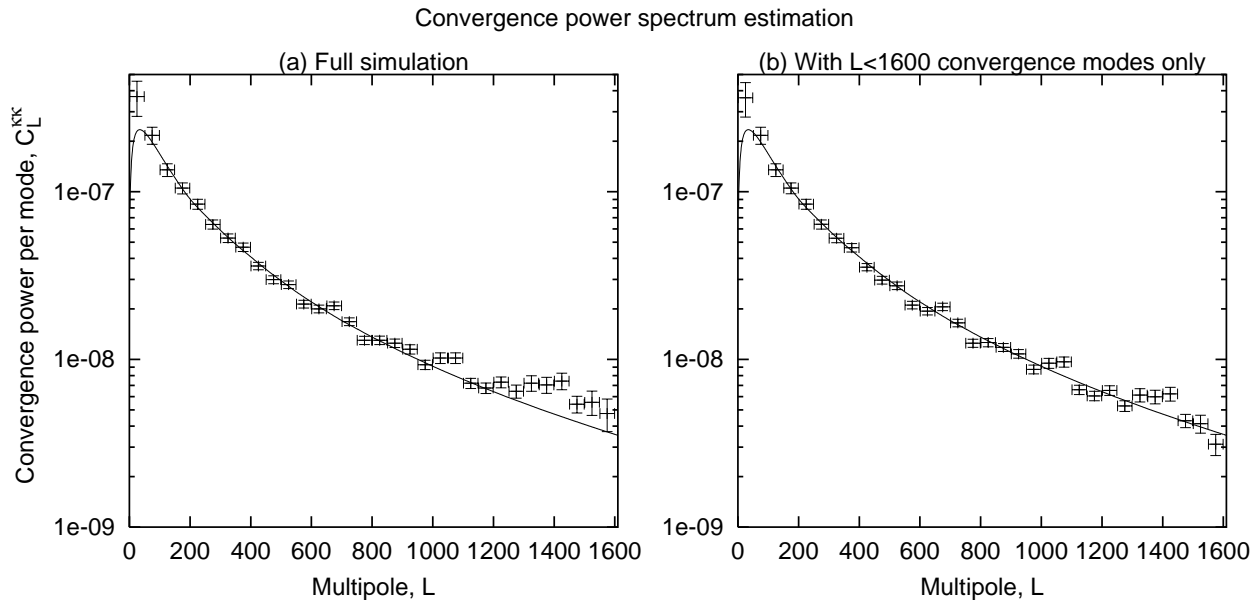}
\caption{\label{fig:clkk}(a) Simulated convergence power spectrum
estimation from Ref. Expt. C with solid angle $0.355$ steradians.  The
solid curve is the fiducial model $C^{\kappa\kappa}_l$; the points are the
convergence power spectrum measured from simulated data after 5
iterations. (b) The same, except that the $l\ge 1600$ convergence modes
were ignored in producing the simulated data (however, exactly the same
CMB+LSS+noise realization was used). The horizontal error bars indicate
the widths of the bins, while the vertical error bars are the $1\sigma$
measurement uncertainties according to Eq. (\ref{eq:gf-sigma}). Note that
the vertical error bars include the Monte Carlo error associated with
using $N_{MC}=3$ simulations to determine $\langle v_\mu\rangle$; if we
had calculated $\langle v_\mu\rangle$ exactly ($N_{MC}\gg 1$), the
vertical error bars would be 13\% smaller (see text for details). }
\end{figure}

In Fig. \ref{fig:clkk}(a), we show a determination of the convergence
power spectrum from simulated data using Ref. Expt. C noise parameters.  
The choice of bins was $N_{bin}=32$, $\Delta l=50$, $l_{\rm max}=1600$,
and the survey area was $0.355$ steradians ($2048$ arcmin $\times$ $2048$
arcmin, with toroidal periodic boundary conditions).  We initialized the
power spectrum estimation with the white spectrum $C^{\kappa\kappa}_l =
1\times 10^{-9}$, corresponding to $c_{\mu,0} = 10^{-9}$. We used
$\zeta_{(p)}=0.5$ for the first two iterations of Eq.
(\ref{eq:poweriter}), which are sufficient to bring the estimated power
spectrum $c_{\mu,n=2}$ to the correct order of magnitude.  Once this
``ballpark'' estimation has been completed, we used $\zeta_{(p)}=1$ for
the subsequent ($n\ge 2$) iterations.

An examination of Fig. \ref{fig:clkk}(a) shows that the power spectrum
estimator Eq. (\ref{eq:poweriter}) has been successful in reproducing the
qualitative features of the power spectrum; however, the power has
evidently been overestimated at the high-$l$ end.  We can perform a
quantitative analysis of the performance of the power spectrum estimator
using the $\chi^2$ test, using the Gaussian error estimate of Eq.
(\ref{eq:gf-sigma}).  The $\chi^2$ value for the $l<1000$ region (where
the power spectrum determination should be cosmic-variance limited) is
$\chi^2=25.46$ for $20$ degrees of freedom ($p=0.18$), indicating that Eq.
(\ref{eq:gf-sigma}) appears to be giving a reasonable estimate of the
uncertainty on the power spectrum.

The same is not true of the high-$l$ region $1000\le l<1600$, for which we
compute $\chi^2=53.05$ for 12 degrees of freedom ($p=4\times 10^{-7}$).  
It is readily apparent from Fig. \ref{fig:clkk} that the failure of the
$\chi^2$ test is due to an upward bias in the power spectrum estimator Eq.
(\ref{eq:poweriter}).  This bias occurs because, regardless of $c_\mu$,
our power spectrum estimator assumes that there is no convergence power at
$l\ge 1600$.  However, it still detects the $B$-modes induced by this
short-wavelength convergence power, and introduces excess convergence
power at $l<1600$ to reproduce these $B$-modes -- hence
$C^{\kappa\kappa}_l$ is overestimated. This bias can be removed by 
either estimating the modes to higher $l$ or including the aliasing of 
power from those modes into the modes we estimate as an additional 
source of noise \cite{1998ApJ...503..492S}. This explanation of the upward bias
is confirmed by Fig. \ref{fig:clkk}(b), in which we have artificially
``turned off'' the lensing effect for convergence modes at $l\ge 1600$,
then produced a simulated data set and applied the power spectrum
estimator Eq. (\ref{eq:poweriter}).  In order to make the comparison
between the original simulation and the restricted (i.e. short-wavelength
lensing turned off) simulation as simple as possible, we have used the
same CMB, noise, and LSS realizations for both.  One can see by comparing
Figs. \ref{fig:clkk}(a) and \ref{fig:clkk}(b) that there is little effect
at low $l$, where the power spectrum estimation is limited by cosmic
variance.  However, at high $l$, one can see that the bias present in the
original simulation has disappeared in the restricted simulation, thereby
confirming that the bias was due to high-$l$ convergence power. The
$\chi^2$ for the $1000\le l<1600$ range has been reduced to $\chi^2=31.36$
(12 degrees of freedom, $p=1.7\times 10^{-3}$), which is still indicative
of underestimation of the uncertainty in the $c_\mu$.  Thus we conclude
that in this regime, either the Gaussian error estimate Eq.
(\ref{eq:gf-sigma}) is underestimating the error by a factor of $\sim
\sqrt{31/12}\approx 1.6$, or the error bars are correlated, or the
iteration of Eq. (\ref{eq:poweriter}) has not completely converged.

In a real lensing experiment, the underlying primary power spectrum
$C^{EE}_l$ is unknown and only the lensed power spectrum is directly
observable (and even our knowledge of this is limited by instrument noise
and cosmic variance).  Thus a slightly more complicated version of the
above analysis will be necessary to simultaneously solve for $C^{EE}_l$
and $C^{\kappa\kappa}_l$.  (Although since in the regime we are examining,
$l<3000$, the $E$ power spectrum is dominated by primary anisotropies
rather than lensing, we do not expect a degeneracy between these two
quantities.)  It will also be necessary to estimate the convergence power
spectrum well beyond the region of interest in order to avoid the upward
bias described here.  Since the signal-to-noise ratio at high $l$ is low,
it will be necessary to use wider bins (i.e. larger $\Delta l$) in this
region.  The choice of exactly which bins to use must be determined by the
characteristics of the specific experiment.

\section{Conclusions}
\label{sec:conclusions}

Weak gravitational lensing of the CMB allows us to reconstruct the
(projected) mass distribution in the universe, thereby probing large-scale
structure and its power spectrum. Since the window functions for lensing
peak at redshift $z$ of order unity, lensing offers the possibility of
using the CMB to study the low-redshift universe
\cite{1999PhRvD..59l3507Z, 1999PhRvL..82.2636S, 2001ApJ...557L..79H,
2001PhRvD..63d3501B, 1997A&A...324...15B}.  ``Cleaning'' of lensing from
CMB maps is potentially valuable for studying the primary CMB,
particularly for inflationary gravitational wave searches using the
low-$l$ $B$-mode polarization \cite{2002PhRvL..89a1303K,
2002PhRvL..89a1304K}. Since the primary CMB polarization is expected to
contain only $E$-modes on the relevant angular scales ($l$ of order
$10^3$), while lensing transfers some of the CMB polarization power into
$B$-modes \cite{1998PhRvD..58b3003Z}, all $B$-modes that we see on these
scales are due to lensing (or foregrounds). Thus the CMB $B$-mode
polarization allows much better lensing reconstruction than is possible
using temperature data alone. It is thus of interest to consider optimal
methods of reconstructing the lensing field from CMB polarization data; in
this paper, we have investigated this problem in detail and improved
significantly on the previous quadratic estimator methods
\cite{2002ApJ...574..566H}. We have shown that this improvement can be up
to an order of magnitude in mean squared error over the zero-noise
reconstruction error for the quadratic estimator.

We make several comments concerning the present calculations.  First of
all, our lensing estimator, Eq. (\ref{eq:wf}), while statistically
superior to the quadratic estimator, still does not achieve the Cramer-Rao
bound on reconstruction accuracy.  We have argued that this results in
part from ``curvature corrections,'' fluctuations in the curvature matrix
that render the Cramer-Rao bound impossible to achieve (more generally,
this should also serve as a warning against blindly assuming that the
statistical errors in any measurement are given by ${\bf F}^{-1}$.). We
expect that our lensing reconstruction estimator is near-optimal since it
is an approximation to the maximum-likelihood estimator and our iterative
estimator shows no signs of incomplete convergence, however the
possibility of further improvement has not been ruled out.

Secondly, we have assumed negligible primary $B$ mode polarization here
(although the formalism described herein is trivially modified to include
significant primary $B$-modes, the results would be qualitatively
different).  In the absence of vector or tensor perturbations, this is
correct; if vector or tensor perturbations are present, then one must
consider their effect on lensing reconstruction.  In the case of
inflationary gravitational waves, primary $B$-modes are generated mostly
on very large angular scales; the arcminute-scale anisotropies used for
lensing reconstruction are uncontaminated \cite{2002ApJ...574..566H}.  
(Formally, if we were doing a lensing reconstruction with the objective of
cleaning lensing contamination of the tensor-induced reionization bump at
$l<20$, we would set $N^{BB}_l=\infty$ for $l<20$ so that the lensing
reconstruction does not remove tensor $B$-modes.)  A more rigorous
investigation of the effect on inflationary gravitational wave searches is
deferred to future work.

Thirdly, the real CMB is contaminated by foregrounds -- an important issue
for all CMB anisotropy experiments. One advantage of using CMB
polarization for lensing reconstruction is that whereas the small-scale
CMB temperature field is heavily contaminated by scattering-induced
secondary anisotropies such as the thermal and kinetic Sunyaev-Zel'dovich
effects, Ostriker-Vishniac effect, and patchy reionization, these effects
are much smaller for polarization \cite{2000ApJ...529...12H}.  However,
polarized point sources and galactic foregrounds are still a serious
concern. These have very different frequency dependence than the blackbody
fluctuations characteristic of the CMB, and this property has been
exploited to remove them; unfortunately, their fluctuation spectrum,
degree of polarization, non-Gaussianity, and variations in frequency
dependence are poorly understood. Galactic foregrounds do not correlate
with the cosmological signals, and in this sense the residuals from their
subtraction
act like instrument noise contaminating the $B$-mode (and, to a lesser
extent, $E$-mode) polarization.
The foreground
power spectrum is likely to be different from that of instrument noise and
is variable across the sky; nevertheless, if the covariance matrix of the
foregrounds (or residuals after foreground subtraction) can be determined,
then we can add the foreground covariance to the instrument noise
covariance matrix ${\bf N}$. (If the statistical properties of
the foreground residuals cannot be determined or at least constrained,
then any cosmological analysis is pointless regardless of the methods
used.)
Polarized point sources produce Poisson
noise; also since many of them are extragalactic, one could be concerned
about their correlation with LSS and hence the lensing signal.  We leave a
detailed study of foregrounds and their impact on lensing reconstruction
to future
investigation. We note that the predicted levels of foreground
contamination from dust and synchrotron galactic emission are at a level
of a few $\mu$K arcmin prior to any frequency cleaning
\cite{2000ApJ...530..133T}, comparable to the noise levels discussed here.  
Frequency cleaning should reduce this, at the expense of amplifying
instrument noise.  If foreground removal is inadequate, this may
result in anomalies in the final results such as unphysical correlations
between the convergence maps and CMB polarization, variation of the
convergence power spectrum between ``clean'' and ``dirty'' portions of the
sky, correlation of the convergence maps with synchrotron or dust
emission, etc.

In summary, we have shown that taking into account the full likelihood
function allows improved reconstruction of the lensing of the CMB
polarization field over that achieved by quadratic statistics.  For
purposes of computing the lensing power spectrum or cross-correlating CMB
lensing with another tracer of the cosmological density field, the most
important improvement is at high $l$ where earlier approaches do not
reconstruct the convergence at high signal-to-noise.  (At low $l$, the
reconstruction is already cosmic variance limited.)  If one's objective is
to clean out the lensing effect in search of primordial gravitational
waves, then the relevant quantity is the residual error in the
reconstruction, and it is important to reduce this even if the convergence
has been mapped at high signal-to-noise; hence improvement at all
multipoles is useful.  We conclude that the likelihood-based estimators
developed here offer the best prospective so far to extract the full
amount of information from future high-resolution CMB polarization
experiments.

\acknowledgments

We would like to thank Wayne Hu and Lyman Page for useful comments. C.H.
acknowledges useful discussions with Mike Kesden, Asantha Cooray, and Marc
Kamionkowski.  C.H. is supported by the NASA Graduate Student Research
Program, grant no. NTGT5-50383.  U.S. acknowledges support from Packard
and Sloan foundations, NASA NAG5-11983 and NSF CAREER-0132953.

\appendix

\section{Quadratic estimator}
\label{sec:hu}

In our simulations, we have compared the error of our iterative estimator,
Eq. (\ref{eq:estcon}), with that of the quadratic estimator, Eq.
(\ref{eq:wfquad}).  Here we show that the latter estimator corresponds to
the optimally weighted quadratic estimator, as proposed by Ref.
\cite{2002ApJ...574..566H}.  Statistically isotropic noise is assumed
throughout.

We begin by expanding Eq. (\ref{eq:qfisher}) using the formula for ${\bf
f}^\kappa_\bfl$ given by Eq. (\ref{eq:fl}).  The off-diagonal elements
vanish by symmetry, while the diagonal elements are:
\begin{equation}
F_l^{\rm (quad)} = {1\over 2}\sum_{\bfl_1} \Tr\left\{
[{\bf f}^\kappa_{-\bfl}]_{-\bfl_1,\bfl_2}
[\langle\hat{\bf C}\rangle_{LSS}]_{l_2}^{-1}
[{\bf f}^\kappa_\bfl]_{\bfl_2,-\bfl_1}
[\langle\hat{\bf C}\rangle_{LSS}]_{l_1}^{-1}\right\},
\label{eq:h1}
\end{equation}
where we have defined $\bfl_2=\bfl-\bfl_1$, and the inverses are $3\times
3$ matrix inverses (using the $\{T,E,B\}$ basis).  Next we note that to
first order in $\kappa$, the correlation between two Fourier modes of
temperature or polarization is:

\begin{equation}
\langle \hat{\bf x}_{\bfl_1} \hat{\bf x}_{-\bfl_2}^\dagger \rangle 
= [{\bf f}^\kappa_\bfl]_{\bfl_1,-\bfl_2} \kappa_\bfl + O(\kappa^2).
\end{equation}
A general quadratic estimator for the convergence $\kappa$ is then
constructed as:
\begin{equation}
\hat\kappa_\bfl = \sum_{\bfl_1} \hat{\bf x}_{-\bfl_2}^\dagger 
[\Xi_{-\bfl}]_{-\bfl_2,\bfl_1} \hat{\bf x}_{\bfl_1},
\label{eq:xi}
\end{equation}
where $\Xi_{-\bfl}$ is the weight matrix, which we assume without loss of
generality to be Hermitian (since the anti-Hermitian part does not
contribute to $\hat\kappa_\bfl$). We further require it to satisfy
$[\Xi_{-\bfl}]_{-\bfl_2,\bfl_1} = [\Xi_\bfl]_{\bfl_2,-\bfl_1}^\dagger$.  
(This guarantees that the estimate $\hat\kappa$ is a real field.)  We can
construct the optimally weighted unbiased (to first order) estimator for
$\kappa$ by minimizing the variance of the estimator (neglecting the
trispectrum contribution):
\begin{equation}
V_l = \langle |\hat\kappa_\bfl|^2\rangle_{LSS}
\approx \sum_{\bfl_1} \Tr\left\{
[\langle\hat{\bf C}\rangle_{LSS}]_{l_2}
[\Xi_{-\bfl}]_{-\bfl_2,\bfl_1}
[\langle\hat{\bf C}\rangle_{LSS}]_{l_1}
[\Xi_\bfl]_{\bfl_1,-\bfl_2} \right\},
\label{eq:v1}
\end{equation}
subject to the constraint that the estimator be unbiased to first-order
(i.e. have unit response):
\begin{equation}
1 = \sum_{\bfl_1} \Tr\left\{ [{\bf f}^\kappa_\bfl]_{\bfl_1,-\bfl_2} 
[\Xi_{-\bfl}]_{-\bfl_2,\bfl_1} \right\}.
\label{eq:v2}
\end{equation}
We may compute the minimum of Eq. (\ref{eq:v1}) constrained by Eq.
(\ref{eq:v2}) using the method of Lagrange multipliers.  The equation
$\delta V_l + \lambda^{-1}\delta 1 = 0$ becomes:

\begin{equation}
\sum_{\bfl_1} \Tr\left\{2[\langle\hat{\bf C}\rangle_{LSS}]_{l_2}
[\delta\Xi_{-\bfl}]_{-\bfl_2,\bfl_1}
[\langle\hat{\bf C}\rangle_{LSS}]_{l_1}[\Xi_\bfl]_{\bfl_1,-\bfl_2}+
\frac{1}{\lambda} [{\bf f}^\kappa_\bfl]_{\bfl_1,-\bfl_2}
[\delta\Xi_{-\bfl}]_{-\bfl_2,\bfl_1} \right\} = 0.
\end{equation}
The solution to this (allowing $\delta\Xi_{-\bfl}$ to be arbitrary) is:
\begin{equation}
[\Xi_{-\bfl}]_{-\bfl_2,\bfl_1} = {1 \over 2\lambda} 
[\langle\hat{\bf C}\rangle_{LSS}]_{l_2}^{-1}
[{\bf f}^\kappa_\bfl]_{\bfl_2,-\bfl_1} 
[\langle\hat{\bf C}\rangle_{LSS}]_{l_1}^{-1}.
\label{eq:xisol}
\end{equation}
The correct normalization $\lambda$ is obtained by substitution into Eq.
(\ref{eq:v2}); it is easily seen to be $\lambda = F_l^{\rm (quad)}$.  The
variance of this estimator in the absence of lensing, determined by
substitution into Eq. (\ref{eq:v1}), is $1/F_l^{\rm (quad)}$.  The
quadratic estimator we have used, Eq. (\ref{eq:wfquad}), is then seen to
be a Wiener-filtered version of Eq. (\ref{eq:xisol}) with the optimized
choice for $\Xi$, Eq. (\ref{eq:xisol}), and its covariance Eq.
(\ref{eq:squad}) then follows from the theory of Wiener-filtering.

Hu and Okamoto \cite{2002ApJ...574..566H} derive a quadratic estimator
using essentially the same method outlined in this appendix.  While they
have chosen to separately optimize the different components of $\Xi$
($TT$, $TE$, $EE$, $TB$, and $EB$) and then combine these to form a
``minimum variance'' estimator, the end result of the optimal filtering
must be the same. (Note that while our covariance response function ${\bf
f}$ is the same as Hu and Okamoto's ${\bf f}$ aside from a factor of
$l^2/2$ due to use of $\Phi$ vs. $\kappa$ as the fundamental field, we
have used $\Xi$ in place of their $F/A$ to avoid confusion with the Fisher
matrix.)

\section{Lensing $B$-modes and idealized reconstruction}
\label{sec:bmode}

The purpose of this Appendix is to investigate the question of whether, in
the absence of noise and field rotation, the equation $B_{\rm unlensed}=0$
could be used to completely reconstruct the convergence field.  We show
that with probability 1, it is possible to reconstruct most of the
convergence modes.  There may remain a small number of convergence modes
that cannot be reconstructed by this method.  If we impose periodic
boundary conditions, the fraction of the convergence modes that are in
this category is at most of order $1/l_{\rm max}$; however, there may be
fewer of these degenerate modes, or possibly none at all. We have not
investigated more realistic survey topologies but we would expect the
general result to be similar on scales small compared to the angular width
of the survey. However, this seems mostly an academic point since zero
noise is of course unrealistic, and there can be many almost-degenerate
modes that spoil a reconstruction based on $B_{\rm unlensed}=0$.

The $B$-mode induced by lensing is, to first order:
\begin{equation}
B_{\bfl,\rm lensing} = {1\over\sqrt{4\pi}} \sum_{\bfl'}
\left(\frac{2}{l'{^2}}\right) \bfl'\cdot(\bfl-\bfl') \sin 2\alpha\;
E_{\bfl-\bfl'} \kappa_{\bfl'} =
\sum_{\bfl'} {\cal T}_{\bfl,\bfl'} \kappa_{\bfl'},
\label{eq:borig}
\end{equation}
where ${\cal T}$ is a transfer matrix that is a linear function of $E$
(and once again $\alpha=\phi_\bfl-\phi_{\bfl'}$).  In the absence of
noise, we may set $B_{\bfl,\rm lensing}$ equal to the observed
polarization $\hat B_\bfl$. We see that if $N$ Fourier modes are
considered, there are $N$ linear equations for $N$ unknowns $\kappa_\bfl$.  
(We do not consider $\bfl=0$ modes since there does not exist a $\kappa_0$
mode, and lensing has no effect on zero-wavenumber CMB modes.)  Thus any
convergence mode that cannot be reconstructed must be associated with a
degenerate direction of ${\cal T}$.  It is clear that for some
realizations of the primary CMB, e.g. $E=0$, ${\cal T}$ is massively
degenerate.  We thus wish to explore whether these singular realizations
are ``likely'' or have probability zero. We will assume here that
$C^{EE}_l>0$ for all of the $E$-modes so that ``probability zero'' and
``measure zero'' can be taken to be equivalent.

In order to do this, we consider the characteristic polynomial of ${\cal
T}$:
\begin{equation}
P(\lambda;E)=\det({\cal T}-\lambda{\bf 1})= \sum_{n=0}^N a_n(E) \lambda^n.
\end{equation}
The determinant of an $N\times N$ matrix is a polynomial of degree $N$ in
the entries of the matrix, hence each $a_n(E)$ is a polynomial of order at
most $N-n$ in the $E_\bfl$.  We know from linear algebra that the roots
and multiplicities of $P(\lambda;E)$ (viewed as a polynomial in $\lambda$)
are precisely the eigenvalues and multiplicities of ${\cal T}$; in
particular, the number of degenerate ($\lambda=0$) modes is equal to the
smallest value of $n$ for which $a_n(E)\neq 0$. Now suppose it were the
case for some $n$ that $a_n(E) \neq 0$ with nonzero probability (recall
that the primary CMB polarization $E$ is a random variable). This implies
that $a_n(E) \neq 0$ is generic, i.e. only a small (measure zero) set of
values of $E$ give $a_n(E)=0$. The significance of this result is that if
we can exhibit even one possible polarization field $E$ for which
$a_n(E)\neq 0$, it follows that $a_n(E)\neq 0$ with unit probability for
the real primary CMB polarization field. A similar statement holds for the
number of degenerate convergence modes: if we can exhibit a possible
polarization field with $n$ degenerate modes, then it follows that with
unit probability, the lensing field as reconstructed from the real CMB
will have at most $n$ degenerate modes.  Conceptually, this means that
the generic lensing reconstruction using $B_{\rm unlensed}=0$ cannot be
more degenerate than any special case we exhibit.  (It may, however, be
less degenerate.)

We consider here the following very simple realization: take a sky with
area $4\pi$, square (with side length $\sqrt{4\pi}$), and with periodic
boundary conditions.  Suppose that only the $E$-mode $E_{\bf L}$ where
${\bf L}=(2\pi/\sqrt{4\pi},0)$ (i.e. the longest-wavelength mode in the
$x$-direction) is nonzero.  Now consider a degenerate convergence mode,
i.e. one that does not contribute to $B_{\rm lensing}$. From Eq.
(\ref{eq:borig}), we see that the $B=0$ requirement forces all of the
convergence modes $\kappa_{\bfl'}$ to be zero except those for which $\sin
2\alpha=0$, i.e. those for which $\bfl'$ is either parallel to or
perpendicular to $\bfl'+{\bf L}$.  The latter is impossible given the
boundary conditions and the former requires $\bfl'$ to lie in the
$x$-direction.  Thus, out of $O(l_{\rm max}^2)$ convergence modes, only
the $O(l_{\rm max})$ modes with wavevector in the $x$-direction cannot be
reconstructed.  Hence no more than a fraction $O(1/l_{\rm max})$ of the
convergence modes are degenerate (i.e. cannot be reconstructed from
$B_{\rm unlensed}=0$), and by the argument of the previous paragraph this
must hold with probability 1 for the actual realization of the primary
polarization field. Note that this is only an upper limit and the 
actual number of degenerate modes may be smaller, or even zero.

The problem of lensing reconstruction using $B_{\rm unlensed}=0$ has been
considered previously using real-space methods by
Ref. \cite{2001PhRvD..63d3501B}.  They derive the following equation for
the lensing-induced $B$-mode:
\begin{equation}
{1\over 2} \nabla^2 B_{\rm lensing}
= \gamma_U \nabla^2 Q - \gamma_Q \nabla^2 U + \nabla \gamma_U
\cdot\nabla Q - \nabla\gamma_Q\cdot\nabla U,
\label{eq:difeq-3}
\end{equation}
where (as above) $\gamma_Q$ and $\gamma_U$ are second derivatives of
$\Phi$.  This is therefore a third-order partial differential equation for
$\Phi$. Ref. \cite{2001PhRvD..63d3501B} then performs a two-dimensional
Taylor-expansion of $\Phi$ and finds that some of the coefficients are not
fixed by Eq. (\ref{eq:difeq-3}).  They thus determine that there exists a
class of lensing potential modes that do not produce $B$-modes purely by
counting the number of equations and the number of variables to be
calculated. The relationship between our approach and that of Ref.
\cite{2001PhRvD..63d3501B} is that we express ${\cal T}$ in the Fourier
basis (Eq. \ref{eq:borig}), whereas they have expressed ${\cal T}$ in the
Taylor polynomial $\{x^jy^k\}_{j,k=0}^\infty$ basis.  These bases are not
equivalent due to the differing boundary conditions assumed (the Fourier
basis imposes periodic boundary conditions whereas the Taylor polynomial
basis does not), and this leads to different conclusions regarding the
number and character of degenerate modes. In the Fourier basis there are
the same number of equations as variables. Thus one cannot conclude from
the counting argument alone that there is a degeneracy in the case of
full-sky coverage. (The analysis on the sphere would be slightly more
intricate since the sky is curved and the boundary conditions have
different topology than in the flat-sky approximation; however, on scales
$l\gg 1$ small compared to the curvature scale, we expect that the results
presented here will still apply.)

\section{Curvature corrections}
\label{sec:a1}

Our purpose in this Appendix is to investigate in greater detail the
mathematical structure of the curvature corrections, i.e. the increase in
uncertainty in lensing reconstruction due to fluctuations of the curvature
matrix.  We show that the curvature correction has another interpretation:
it represents the increased noise in the reconstruction of $\kappa_\bfl$
due to the presence of other lensing modes, $\kappa_{\bfl'}$ (where
$\bfl'\neq\pm\bfl$), an effect studied in detail in Kesden {\slshape et
al} \cite{2003astro.ph..2536K}, where it was called the ``first-order
noise contribution'' and denoted by $N^{(1)}_{XX',X''X'''}(L)$. Here we
show that in fact the curvature correction contains the likelihood
analysis manifestation of the $N^{(1)}$ of Ref.
\cite{2003astro.ph..2536K}.

We can compute the curvature corrections to this as follows.  If we define
$\delta{\cal F} = {\cal F} - {\bf F}$, and retain our approximation from
Eq. (\ref{eq:fphi}) that ${\bf F}$ is independent of $\kappa$, then we
have:
\begin{equation}
{\bf S} = \langle (\delta{\cal F} + {\bf F} 
+ {\bf C}^{\kappa\kappa\,-1})^{-1} \rangle_{LSS}
\sim {\bf S}_0 + {\bf S}_0 \langle \delta{\cal F} \; {\bf
S}_0\; \delta{\cal F} \rangle_{LSS} {\bf S}_0 - ...,
\label{eq:cc}
\end{equation}
where ${\bf S}_0 = ({\bf F} + {\bf C}^{\kappa\kappa\,-1})^{-1}$.  [To
derive this equation, we have merely Taylor-expanded in $\delta{\cal F}$,
then taken the CMB+noise+LSS ensemble average, and noted that by
definition $\delta{\cal F}$ vanishes when ensemble-averaged over CMB+noise
realizations. Note that because we have taken the expectation value, Eq.
(\ref{eq:cc}) should be viewed as an asymptotic expansion rather than a
Taylor expansion.] The mean squared error ${\bf S}$ picks up ``curvature
correction'' terms involving $\delta{\cal F}$ that cause it to not equal
the naive result ${\bf S}_0$.  Note that curvature corrections to ${\bf
S}$ only increase ${\bf S}$, they cannot decrease it (in the sense that
${\bf S}-{\bf S}_0$ has all eigenvalues $\ge 0$; equivalently the diagonal
elements ${\bf S}_{jj}\ge[{\bf S}_0]_{jj}$ in all orthonormal bases).  
This is true to all orders in $\delta{\cal F}$ because the inverse of the
mean of a set of positive definite Hermitian matrices is smaller than the
mean of the inverses (in this same sense).

In order to compute the second-order curvature correction explicitly, we
must understand the fluctuations in the curvature matrix.  For simplicity,
we evaluate $\delta{\cal F}$ at $\kappa=0$.  In this case, we find:
\begin{eqnarray}
{\cal F}_{\bfl,\bfl'} \equiv && {\cal F}[\kappa_\bfl,\kappa_{\bfl'}] =
{\partial^2\over\partial\kappa_\bfl^\ast\partial\kappa_{\bfl'}}
\left( {1\over 2}\ln\det\hat{\bf C}_g + {1\over 2}\hat{\bf
x}^\dagger\hat{\bf C}_g^{-1}{\bf x} \right)
\nonumber \\
= &&
{1\over 2}\hat{\bf x}^\dagger \hat{\bf C}^{-1} 
\left( 2 {\partial\hat{\bf C}\over\partial \kappa_{-\bfl}}
\hat{\bf C}^{-1} {\partial\hat{\bf C}\over\partial \kappa_{\bfl'} } -
{\partial^2\hat{\bf C}\over\partial \kappa_{-\bfl} \partial \kappa_{\bfl'}} 
\right) \hat{\bf C}^{-1}\hat{\bf x}
-{1\over 2} \Tr\left[ \hat{\bf C}^{-1} \left( {\partial\hat{\bf
C}\over\partial \kappa_{-\bfl}}
\hat{\bf C}^{-1} {\partial\hat{\bf C}\over\partial \kappa_{\bfl'} } -
{\partial^2\hat{\bf C}\over\partial \kappa_{-\bfl} \partial \kappa_{\bfl'}} 
\right)\right],
\end{eqnarray}
where the second line has the $\hat{\bf C}$'s evaluated at $\kappa=0$.  
Subtracting out the average value ${\bf F}_{\bfl,\bfl'}$ over CMB+noise
realizations yields:
\begin{equation}
\delta{\cal F}_{\bfl,\bfl'} 
= -{1\over 2}\hat{\bf x}^\dagger \hat{\bf C}^{-1} {\bf J}_{(-\bfl,\bfl')}
\hat{\bf C}^{-1} \hat{\bf x}
+ {1\over 2}\Tr\left(\hat{\bf C}^{-1} {\bf J}_{(-\bfl,\bfl')}\right),
\end{equation}
where we have defined:
\begin{equation}
{\bf J}_{(-\bfl,\bfl')} = \left.\hat{\bf C} 
{\partial^2[\hat{\bf C}^{-1}]\over \partial\kappa_{-\bfl}
\partial\kappa_{\bfl'}} \hat{\bf C} \right|_{\kappa=0}
= \left.\left( {\partial\hat{\bf C}\over\partial \kappa_{-\bfl}}
\hat{\bf C}^{-1} {\partial\hat{\bf C}\over\partial \kappa_{\bfl'} } +
{\partial\hat{\bf C}\over\partial \kappa_{\bfl'}}
\hat{\bf C}^{-1} {\partial\hat{\bf C}\over\partial \kappa_{-\bfl} } -
{\partial^2\hat{\bf C}\over\partial \kappa_{-\bfl} \partial
\kappa_{\bfl'}} \right) \right|_{\kappa=0}.
\label{eq:j}
\end{equation}
Note that ${\bf J}_{(-\bfl,\bfl')} = {\bf J}_{(\bfl',-\bfl)} = {\bf
J}_{(\bfl,-\bfl')}^\dagger$.

Using this relation, and noting that at $\kappa=0$ we have
$\partial\hat{\bf C}/\partial\kappa_\bfl = {\bf f}^{\kappa}_\bfl$, we can
use Wick's theorem to compute:
\begin{equation}
\langle \delta{\cal F}_{\bfl,\bfl'} \delta{\cal F}_{\bfl_1,\bfl'_1} \rangle =
{1\over 4} \Tr\left( \hat{\bf C}^{-1} {\bf J}_{(-\bfl,\bfl')}
\hat{\bf C}^{-1} {\bf J}_{(-\bfl',\bfl)} \right).
\label{eq:trace4}
\end{equation}
In the case of statistically isotropic noise, Eq. (\ref{eq:trace4}) allows
us to compute the covariance of the reconstruction ${\bf S}$ using Eq.
(\ref{eq:cc}).  In harmonic space, the off-diagonal elements of ${\bf S}$
vanish by symmetry whereas the diagonal elements are given by:

\begin{equation}
S_l = [S_0]_l + {1\over 4} [S_0]_l^2 \sum_{\bfl'}  [S_0]_{l'}\Tr\left(
\hat{\bf C}^{-1} {\bf J}_{(-\bfl,\bfl')}
\hat{\bf C}^{-1} {\bf J}_{(-\bfl',\bfl)} \right),
\label{eq:s2}
\end{equation}
which is a summation over quadrilateral configurations of the modes
$\bfl$, $\bfl'$, and the mode over which we sum when computing the trace.

To lowest order ($\delta{\cal F}^2$), the curvature correction is given by
Eq. (\ref{eq:s2}). The correction to the mean inverse curvature $V_l$,
i.e. to the covariance matrix of an unbiased estimator for $\kappa$, is
related to the correction to $S_l$ by noting that $S_l^{-1} = V_l^{-1} +
C_l^{\kappa\kappa\,-1}$, hence:
\begin{equation}
\Delta V_l = \frac{V_l^2}{S_l^2} \Delta S_l \approx
{1\over 4} V_l^2 \sum_{\bfl'}  [S_0]_{l'} \Tr\left(
\hat{\bf C}^{-1} {\bf J}_{(-\bfl,\bfl')}
\hat{\bf C}^{-1} {\bf J}_{(-\bfl',\bfl)} \right).
\label{eq:deltav}
\end{equation}
We now pass to the ``linear approximation'' in which the second derivative
$\partial^2\hat{\bf C}/\partial\kappa_\bfl^\ast\partial\kappa_{\bfl'}$ is
neglected.  (This was found to be a valid approximation for
temperature-based lensing estimation on scales $l<3500$
\cite{2003PhRvD..67d3001H}, although it is unclear whether this is also
true in the present context.)  Substituting Eq. (\ref{eq:j}) then yields:

\begin{equation}
\Delta V_l \approx {1\over 2} V_l^2 \sum_{\bfl'}  [S_0]_{l'}
\Tr\left(  \hat{\bf C}^{-1} {\bf f}_{\bfl}^\kappa
\hat{\bf C}^{-1} {\bf f}_{\bfl'}^\kappa
\hat{\bf C}^{-1} {\bf f}_{-\bfl}^\kappa
\hat{\bf C}^{-1} {\bf f}_{-\bfl'}^\kappa  +
\hat{\bf C}^{-1} {\bf f}_{\bfl}^\kappa
\hat{\bf C}^{-1} {\bf f}_{-\bfl}^\kappa
\hat{\bf C}^{-1} {\bf f}_{\bfl'}^\kappa
\hat{\bf C}^{-1} {\bf f}_{-\bfl'}^\kappa  \right).
\label{eq:deltav2}
\end{equation}

This should be compared with the first-order noise contribution $N^{(1)}$
of Ref. \cite{2003astro.ph..2536K}.  In our notation, and written in terms
of the convergence rather than the potential, their Eq. (25) can be
re-written with the help of some algebra and the relation ${\bf
f}^\kappa_\bfl={\bf f}_{-\bfl}^{\kappa\,\dagger}$ as:

\begin{equation}
N^{(1)}_{TT,TT}(\kappa_\bfl) 
= {1\over 2}V_l^2 \sum_{\bfl'} C^{\kappa\kappa}_{l'} \sum_{\bfl_1} 
\frac{  [f_\bfl]_{\bfl_1,-\bfl_2}^\kappa
[f_{\bfl'}]_{-\bfl_2,-\bfl_2-\bfl'}^\kappa
[f_{-\bfl}]_{-\bfl_2-\bfl',\bfl_1-\bfl'}^\kappa
[f_{-\bfl'}]_{\bfl_1-\bfl',\bfl_1}^\kappa   }{
[\langle\hat{\bf C}\rangle_{LSS}]_{l_1}^\kappa
[\langle\hat{\bf C}\rangle_{LSS}]_{l_2}^\kappa
[\langle\hat{\bf C}\rangle_{LSS}]_{\bfl_2+\bfl'}^\kappa
[\langle\hat{\bf C}\rangle_{LSS}]_{\bfl_1-\bfl'}^\kappa }.
\end{equation}
This is the first term of Eq. (\ref{eq:deltav2}), except that it only
includes temperature information (hence we have multiplication of numbers
rather than $3\times 3$ matrices), and the residual power spectrum
$[S_0]_{l'}$ has been replaced with the raw convergence power spectrum
$C^{\kappa\kappa}_{l'}$. (The latter difference arises because Ref.
\cite{2003astro.ph..2536K} computed the first-order noise $N^{(1)}$ for
the quadratic estimator; when the Bayesian estimator is used, the
contaminating modes $\{\kappa_{\bfl'}\}$ have their power reduced from
$C^{\kappa\kappa}_{l'}$ to $[S_0]_{l'}$ since the estimated lensing field
$\hat\kappa$ is used to de-lens the CMB.)  Thus we see that the
first-order noise arises in the likelihood formalism as a curvature
correction, which is not taken into account in the Fisher matrix for
lensing reconstruction.

The question naturally arises as to the interpretation of the second term
in the curvature correction, Eq. (\ref{eq:deltav2}).  We note that within
the linear approximation,
\begin{equation}
\langle\hat{\bf C}^{-1}\rangle_{LSS} 
= \hat{\bf C}_{(0,0)}^{-1} 
+ \sum_{\bfl'} [S_0]_{\bfl'} \hat{\bf C}_{(0,0)}^{-1}
{\bf f}^\kappa_{\bfl'} \hat{\bf C}_{(0,0)}^{-1} {\bf f}_{-\bfl'}^\kappa
\hat{\bf C}_{(0,0)}^{-1} .
\end{equation}
The second term of Eq. (\ref{eq:deltav2}) is thus seen to be the
correction to the Fisher matrix, Eq. (\ref{eq:fphi}), due to the lensing
effect on the CMB power spectrum.

\end{document}